\documentclass{jfm}

\usepackage{graphicx}
\usepackage{newtxtext}
\usepackage{newtxmath}
\usepackage{natbib}
\usepackage{hyperref}
\usepackage{caption}
\usepackage{xcolor}
\usepackage{scalerel}
\usepackage{multirow}
\usepackage{multicol}
\usepackage{booktabs}
\usepackage{subfig}
\newcommand{\ra}[1]{\renewcommand{\arraystretch}{#1}}
\hypersetup{
    colorlinks = true,
    urlcolor   = blue,
    citecolor  = black,
}

\newcommand{\RomanNumeralCaps}[1]

\newcommand{\R}[1]{\textcolor{black}{#1}}

\title{On the structure and dynamics of secondary flows over multi-column roughness in channel flow}

\author{A. S. Sathe\aff{1},
  W. Anderson\aff{2},
  M. Calaf\aff{3} and
  M. G. Giometto\aff{1}
  \corresp{\email{mg3929@columbia.edu}}}

\affiliation{\aff{1}Department of Civil Engineering and Engineering Mechanics, Columbia University, New York, NY 10027
\aff{2}Department of Mechanical Engineering, The University of Texas at Dallas, 800 West Campbell Road,
Richardson, TX 75080, USA
\aff{3}Department of Mechanical Engineering, University of Utah, Salt Lake City, Utah 84112, USA}

\begin{document}
\maketitle
\makeatletter
\renewcommand{\absfooterflag}{}%
\renewcommand{\pagelimitfooter}{}%
\makeatother

\begin{abstract}
Secondary flows induced by spanwise heterogeneous surface roughness play a crucial role in determining engineering-relevant metrics such as surface drag, convective heat transfer, and the transport of airborne scalars.
While much of the existing literature has focused on idealized configurations with regularly spaced roughness elements, real-world surfaces often feature irregularities, clustering, and topographic complexity for which the secondary flow response remains poorly understood.
\R{Motivated by this gap,} we investigate multi-column roughness configurations that serve as a regularized analog of roughness clustering.
\R{Using large-eddy simulations, we systematically examine secondary flows across a controlled set of configurations in which cluster density and local arrangement are varied in an idealized manner, and observe that these variations give rise to distinct secondary flow polarities.}
Through a focused parameter study, we identify the spanwise gap between the edge-most roughness elements of adjacent columns, normalized by the channel half-height ($s_a/H$), as a key geometric factor governing this polarity.
In addition to analyzing the time-averaged structure, we investigate how variations in polarity affect the instantaneous dynamics of secondary flows.
Here, we find that the regions of high- and low-momentum fluid created by the secondary flows alternate in a chaotic, non-periodic manner over time.
Further analysis of the vertical velocity signal shows that variability in vertical momentum transport is a persistent and intrinsic feature of secondary flow dynamics.
Taken together, these findings provide a comprehensive picture of how the geometric arrangement of roughness elements governs both the mean structure and temporal behavior of secondary flows.
\end{abstract}

\vspace{1ex}
\noindent\rule[4pt]{\textwidth}{0.4pt} 
\vspace{2ex}

\section{Introduction} \label{sec:intro}

Understanding the dynamics of turbulent boundary layers (TBLs) is a long-standing challenge in fluid mechanics, with foundational studies dating back a hundred years \citep{Prandtl1925, Nikuradse1933, Colebrook, Schlichting}.
TBLs occur over various surfaces, including urban environments \citep{oke_mills_christen_voogt_2017, Oke1982, Meili2020}, wind farms arrays \citep{Calaf2010, Meyers2013, Cortina2020}, ships \citep{Grigson, Schultz2000} and turbomachinery \citep{Acharya} to name a few.
When these rough surfaces exhibit spanwise heterogeneity comparable to the boundary layer height, they give rise to roughness-induced secondary flows.
Unlike secondary flows of the first kind, which arise from vortex stretching and tilting mechanisms typically induced by curvature or rotation in the mean flow, these are the secondary flows of the second kind which arise from gradients in Reynolds stresses and are maintained through anisotropic turbulent production and dissipation \citep{Hinze1967, Anderson2015Secflow}. 
These secondary flows, characterized by streamwise vortical structures that redistribute momentum and energy in the spanwise-wall-normal plane, can significantly modify the mean flow field, alter turbulence intensities, and influence quantities of engineering relevance such as surface drag and convective heat transfer.

Over the past decade, substantial research efforts have been devoted to advancing our understanding of secondary flows induced by surface roughness heterogeneities. 
Numerous experimental investigations \citep{Barros2014, Vanderwel2015, Kevin2017, Medjnoun2020, Wangsawijaya2020, Womack2022} and numerical studies \citep{Willingham2014, Anderson2015Secflow, Yang2018_updated, Hwang2018, Stroh2020, Joshi2022, Kaminaris2023, Sathe2024} have demonstrated the prevalence and significance of these flows across various roughness configurations. 
Typically, these secondary flows manifest as pairs of counter-rotating vortical structures whose dimensions scale with the characteristic spanwise spacing of the roughness elements. 
A distinctive hallmark of these secondary motions is the spanwise undulation of the mean streamwise velocity throughout the flow depth, resulting in clearly delineated low- and high-momentum pathways (commonly termed LMPs and HMPs, respectively) \citep{Barros2014}. 
Specifically, LMPs are bounded by vortical pairs oriented such that these pathways align with regions exhibiting upward vertical velocities. 
In contrast, HMPs align with regions of downward vertical velocities. 
These alternating momentum pathways recur periodically along the spanwise direction, directly reflecting the periodic arrangement of the roughness heterogeneity.

The self-sustaining nature of roughness-induced secondary flows is fundamentally tied to spanwise heterogeneities in the Reynolds stress components, which actively generate and maintain streamwise vorticity through turbulence anisotropy and vorticity production mechanisms \citep{Anderson2015Secflow}. 
Gradients in both Reynolds normal and shear stresses arise due to spatial variations in surface roughness, which modulate the local turbulence intensity and stress distributions across the span. 
An alternative yet complementary explanation for the maintenance of these secondary motions was proposed by \citet{Hinze1967}, based on the energy budget of turbulent kinetic energy (\textit{tke}). 
Assuming streamwise homogeneity and negligible transport contributions—assumptions valid for strip-type roughness configurations—the \textit{tke} budget reduces to a balance between production, dissipation, and advection. 
In this framework, local imbalances between production ($\mathcal{P}$) and dissipation ($\epsilon$) necessitate advection of \textit{tke} in the spanwise-wall-normal plane. 
Regions of excess production draw in turbulence-deficient fluid from the outer layer, inducing a downdraft above high-production zones.
This fluid, having acquired elevated turbulence levels due to intense local shear and Reynolds stress production, becomes turbulence-rich and is laterally transported toward regions dominated by dissipation, generating spanwise outflow near the wall.
The resulting turbulence-deficient fluid is advected upward, closing the circulation and giving rise to spanwise-periodic roll cells with distinct zones of updraft and downdraft. 
This cyclic redistribution of turbulence and momentum establishes that secondary flows are sustained by the spanwise modulation of $\mathcal{P} - \epsilon$ imposed by the underlying surface heterogeneity.

Analysis of flow over spanwise heterogeneous strip-type roughness has revealed that regions of elevated turbulent production coincide with areas of high roughness, whereas regions of excess dissipation align with low-roughness regions \citep{Anderson2015Secflow}.
This spatial disparity in the turbulence budget gives rise to secondary circulations in which LMPs and their associated updrafts emerge over low-roughness regions, while HMPs and corresponding downdrafts align with high-roughness areas.
A similar alignment was reported in the experimental study of \citet{Barros2014}, where flow over a turbine blade—damaged by the deposition of foreign material—exhibited LMPs over recessed roughness and HMPs over elevated deposits.
In contrast, \citet{Vanderwel2015}, in their experiments involving streamwise-aligned LEGO blocks with varying spanwise spacing (commonly termed ``ridge-type" roughness), observed the opposite: LMPs were found over the elevated roughness, while HMPs occupied the recessed regions.
This reversal in the alignment of secondary flow features—often referred to as flow polarity—has been a subject of ongoing debate ever since.

Direct numerical simulations by \citet{Hwang2018} of spatially developing turbulent boundary layer over smooth ridges also showed LMPs aligned with high roughness and HMPs with low roughness.
In their analysis of the \textit{tke} budget equation, they uncovered that, similar to strip-type roughness, the production values exceed the dissipation value over the ridges.
While their analysis of the \textit{tke} budget showed $\mathcal{P} - \epsilon > 0$ over the ridges, consistent with earlier findings from strip-type roughness, they found that the transport term ($T$) in the energy equation plays a critical role for ridge-type geometries and cannot be neglected. 
When included, the transport term reverses the net imbalance, such that $\mathcal{P} - \epsilon - T < 0$, implying that an upward advection of \textit{tke} is necessary over the ridges—thereby flipping the polarity of the secondary flows compared to the strip-type roughness. 
These findings firmly establish that either LMPs or HMPs (and their associated vertical motions) can align with the regions of high or elevated roughness, and that the polarity is dictated by the specific arrangement and geometry of the roughness elements.

Following the emergence of conflicting observations regarding the alignment of secondary flow structures with regions of high roughness, identifying the rough-surface parameters that influence flow polarity has become a central topic in the secondary flow community.
\citet{Yang2018_updated} investigated flow over streamwise-aligned rows of pyramidal obstacles and found that the lateral spacing between roughness features, $s_y$, when normalized by the boundary layer height $H$, acts as a governing roughness parameter.
Their results showed that for $s_y/H \lesssim 1$, updrafts tend to prevail over the pyramids—suggesting a flow polarity analogous to ridge-type roughness—while for $s_y/H \gtrsim 1$, domain-scale mean circulations emerged, with downdrafts aligning over the roughness rows-suggesting a flow polarity analogous to strip-type roughness.
Building on the broader effort to characterize polarity-controlling parameters, \citet{Stroh2020} conducted DNS of flow over streamwise-aligned roughness strips and identified the relative elevation between smooth and rough regions as a key determinant of secondary flow polarity. 
When the smooth surface was recessed below the rough elements, updrafts formed over the high-roughness zones.
However, raising the smooth surface to twice the roughness height reversed the polarity, with downdrafts now appearing over the rough strips. 
An intermediate state was observed when the smooth and rough surfaces were at the same height. 
Their analysis linked these changes in secondary-flow polarity to wall-normal deflections of spanwise velocity fluctuations, reflected in the sign change of the spanwise–wall-normal Reynolds stress component.
Further contributions from \citet{Joshi2022}, based on LES of flow over streamwise-aligned rows of synthetic trees, emphasized the importance of streamwise spacing between roughness elements. 
As the configuration transitioned from d-type to k-type roughness with increasing streamwise gap, they observed a corresponding reversal in secondary flow polarity—from updrafts over roughness rows (d-type) to downdrafts over them (k-type). 
Collectively, these studies establish that lateral spacing of spanwise heterogeneities, relative surface elevations, and streamwise roughness spacing are critical parameters in determining secondary flow polarity for a given roughness configuration.

Another layer of complexity in understanding secondary flows lies in their temporal behavior. 
While long-time-averaged fields often depict coherent, domain-scale counter-rotating vortices, recent studies have revealed that these structures are, in fact, the statistical imprint of smaller, intermittent, and dynamically evolving motions. 
\citet{Kevin2017} investigated flow over a converging–diverging riblet pattern and demonstrated that the instantaneous secondary motions are composed of strong, compact vortices, far more intense than the large-scale structures observed in the mean flow.
Their conditional averaging revealed that only 28\% of the snapshots bore resemblance to the pair of counter-rotating vortices seen in long-time-averaged fields. Interestingly, in 31\% of the cases, the flow exhibited large-scale, one-sided rotational motions, which occurred more frequently than symmetric vortex pairs. The remaining 41\% of the snapshots showed no resemblance to the time-averaged secondary flow structures.
This lead to the conclusion that the domain-scale secondary flows are artifacts of time-averaging, obscuring a highly transient flow field beneath.
This view was refined by \citet{Vanderwel2019}, who, through a combination of experiments and DNS over ridge-type roughness, revealed that the large-scale secondary flows observed in the time-averaged fields are the ensemble imprint of numerous compact, counter-rotating vortex pairs appearing in the instantaneous flow. 
These structures, smaller in size yet significantly stronger than the mean vortices, occur in varying orientations and are distributed asymmetrically across the span. 
Their findings emphasized that the apparent coherence of secondary flows in the mean arises from a non-homogeneous spatial distribution of these transient vortical events, rather than from persistent, rigid roll structures.
Additionally, \citet{Anderson2019} modeled the secondary flow system using a reduced-order dynamical framework and demonstrated the potential for spontaneous polarity reversals, indicating chaotic, non-periodic behavior in the evolution of these flows akin to chaotic attractors.
Additional insights into the temporal dynamics of secondary flows were provided by \citet{Wangsawijaya2020}, who showed that the secondary flows meander and shift in response to spanwise heterogeneity, with time-dependence most pronounced when the roughness wavelength matches the boundary layer thickness ($s_y/H \approx 1$). 
Similar to the aforementioned studies, their results emphasized that the apparent coherence of secondary flows may be a manifestation of averaging over time-dependent, spatially localized motions. 
Collectively, these studies indicate that secondary flows are not steady-state features but evolve through a complex interplay of intermittent vortices, unsteady momentum pathways, and potentially chaotic dynamics, all of which must be considered for a complete understanding of their behavior.

Recently, \citet{Womack2022} reported the existence of secondary flows over a surface populated with irregularly arranged truncated cones, despite the absence of any regularized spanwise heterogeneity in the roughness layout.
They hypothesized that these secondary motions originate near the leading edges of the roughness topography and can persist over extended streamwise distances, even as the downstream topography varies significantly.
This hypothesis was subsequently examined and supported through detailed DNS investigations by \citet{Kaminaris2023}.
However, in their analysis, the roughness configurations were deliberately designed to eliminate clustering and directional bias, focusing instead on more homogeneously distributed roughness. 
While the study of such surfaces is certainly relevant, clustering and directional bias are also common features in many real-world environments. 
Examples include the patchy deposition of foreign materials on turbine blades
\citep{Barros2014}, marine biofouling on ship hulls and offshore structures \citep{Schultz2007}, random ice accretion on aircraft surfaces \citep{BRAGG2005}, and atmospheric boundary layers over complex terrain \citep{Giometto2016}. 
This raises an important question: how might the presence of roughness clustering, where multiple closely spaced elements collectively act as a high-drag region, influence the structure and dynamics of secondary flows?
To date, most studies have focused on canonical configurations where each streamwise-aligned high-drag zone typically corresponds to a single roughness element in the spanwise direction \citep{Vanderwel2015, Yang2018_updated, Hwang2018, Joshi2022, Sathe2024}.
Given the critical role the secondary flows play in determining engineering-relevant metrics such as surface drag, heat transfer, and momentum redistribution, it is essential to understand how parameters such as cluster density and local topography influence these resulting circulations.

\R{To advance this understanding}, we investigate a multi-column roughness configuration, visualized in figure~\ref{fig:wt_config}, that serves as a regularized analog of roughness clustering. 
This setup allows for precise control of geometric parameters, enabling systematic exploration of secondary flow polarity. 
The configuration consists of multiple resolved roughness elements combined in both the streamwise and spanwise directions to form well-defined, streamwise-aligned high-roughness regions.
Here, individual roughness elements are modeled as momentum-absorbing obstacles resembling wind turbines, providing a simplified representation of flow through porous media \citep{Calaf2010}.
We vary cluster density by systematically adjusting the lateral spacing between roughness elements within each column, while maintaining a constant streamwise gap across all cases. 
In addition, we examine the impact of local arrangement by comparing staggered and aligned configurations within the multi-column setup.
Overall, this study aims to advance our understanding of how roughness clustering influences secondary flow behavior by examining both time-averaged and instantaneous flow characteristics within this controlled, yet physically relevant framework.

The paper is organized as follows: 
Section~\ref{sec:methodology} outlines the methodology employed in this study, including the simulation algorithm (\S\ref{sec:algo}) and the simulation setup, along with the naming conventions used to describe the various configurations (\S\ref{sec:setup}).
The characteristics of the time-averaged secondary flows are examined in section~\ref{sec: Time averaged results}, where we identify a key parameter that governs the polarity of secondary flows in the multi-column roughness configuration. 
Section~\ref{sec: Instantaneous results} explores the temporal dynamics of secondary flows with differing polarities, providing insight into their unsteady behavior.
Finally, section~\ref{sec: Conclusion} provides the conclusions drawn from the study.

\section{Methodology } \label{sec:methodology}

\subsection{Simulation algorithm} \label{sec:algo}

A suite of LES of flow over windfarm arrays is performed in this study using an in-house code \citep{albertson1999natural,albertson1999surface, bou2005scale,chamecki2009large, Calaf2010, Meyers2013, anderson2015numerical, fang2015large, li2016quality, Giometto2016, Sharma2016, Cortina2020, Li_Giometto_2024, Hora2024}.
The filtered Navier-Stokes equations are solved in their rotational form ~\citep{orszag1975numerical} to ensure the conservation of mass and kinetic energy in the inviscid limit, i.e.,
\begin{equation}
    \frac{\partial \tilde{u}_i}{\partial x_i}=0\ ,
    \label{eq:continuity}
\end{equation}
\begin{equation}
    \frac{\partial \tilde{u}_i}{\partial t} +  \tilde{u}_j (\frac{\partial \tilde{u}_i}{\partial x_j}-\frac{\partial \tilde{u}_j}{\partial x_i})  = - \frac{1}{\rho} \frac{\partial \tilde{p}^*}{ \partial  x_i} - \frac{\partial \tau_{ij}^{SGS}}{\partial x_j} - \frac{1}{\rho }\frac{\partial \tilde{p}_\infty}{ \partial x_1} \delta_{i1}+\tilde{F}_i\ ,
    \label{eq:momentum}
\end{equation}
where $\tilde{u}_1$, $\tilde{u}_2$, and $\tilde{u}_3$ are the filtered velocities along the streamwise $x_1$, spanwise $x_2$, and wall-normal $x_3$ directions, respectively and $\rho$ is the reference density.
In this study, index notation is used for compact representation of governing equations, while the more conventional meteorological notation—where $u$, $v$, and $w$ denote the velocities in streamwise ($x$), spanwise ($y$) and vertical ($z$) direction respectively—is used in the discussions.
The deviatoric component of the subgrid-scale (SGS) stress tensor ($\tau_{ij}^{SGS}$) is evaluated via the Lagrangian scale-dependent dynamic (LASD) Smagorinsky model \citep{bou2005scale}.
The LASD model has undergone rigorous validation in a range of turbulent flow scenarios, including wall-modeled large-eddy simulations of the atmospheric boundary layer \citep{momen2017mean, salesky2017nature}, as well as in complex flows over urban-like surfaces and canopy geometries \citep{anderson2015numerical, li2016quality, Giometto2016, yang2016mean, Li_Giometto_2023, Sathe2024}.
Given the high Reynolds number regime considered, viscous stresses are omitted, and wall stress is instead computed using an inviscid logarithmic law of the wall appropriate for fully rough surfaces \citep{Giometto2016}.
Neglecting viscous stresses is valid under the assumption that SGS stress contributions are predominantly from the pressure field.
To incorporate SGS pressure effects and resolved turbulent kinetic energy (\textit{tke}), the pressure field is redefined using a modified form: $\tilde{p}^*=\tilde{p}+\frac{1}{3}\rho \tau_{ii}^{SGS}+\frac{1}{2} \rho \tilde{u}_i \tilde{u}_i$.
The flow is driven by a spatially uniform pressure gradient.
The magnitude of friction velocity $u_\tau$ is linked to this pressure gradient through the relation $(-\nabla p/\rho) H = u^2_\tau$, where $H$ is the half channel height.
This allows the friction velocity to be an input parameter for this study. 
Periodic boundary conditions are imposed in the streamwise and spanwise directions, while the upper boundary has a free-slip boundary condition, satisfying $w=0, \partial u/ \partial z=0 \ \text{and} \ \partial v/ \partial z=0$.
The lower surface consists of arrays of wind turbines, where the standard actuator disk model without rotation is used to parameterize the turbine induced forces ($\tilde{F}_i$) \citep{Jimenez2007, Calaf2010, Wu2011}.
Based on the scope of this study, the actuator disk model with rotation is not introduced as the model without rotation can accurately capture first- and second-order statistics in the far-wake region \citep{Wu2011} as well as the mean kinetic energy pathways \citep{Meyers2013}.
It is important to note that this is not a wind energy study; rather, wind turbines are used here as representative porous roughness elements commonly found in both built and natural environments.

The spatial derivatives in the wall-parallel directions are performed using a pseudo-spectral collocation method that relies on truncated Fourier expansions \citep{orszag1970analytical}.
In contrast, discretization in the vertical direction employs a second-order accurate finite-difference scheme on a staggered grid.
Time advancement is carried out using an explicit second-order Adams–Bashforth scheme. 
To accurately handle the non-linear advection terms and prevent aliasing errors, a 3/2 padding rule is applied \citep{canuto2007spectral, margairaz2018comparison}. 
The incompressibility constraint, expressed in equation (\ref{eq:continuity}), is enforced through a fractional-step method \citep{kim1985application}.
The simulations are averaged for $120T$, where $T$ is the large eddy turnover time defined as $T = H/u_\tau$ to ensure temporal convergence of first- and second-order statistics.
This averaging period is about $7\times$ larger than what is typically used in simulations of flow over obstacles \citep{Coceal2006, Claus2012, Tian2021}.
The time step employed in these simulations is selected to maintain a Courant-Friedrichs-Lewy (CFL) number below 0.1, ensuring numerical stability.

\subsection{Simulation setup and naming convention} \label{sec:setup}
Different roughness element arrangements are explored in this study to analyze the secondary flow response in a multi-column setup.
The size of the computational domain is $[0, L_x] \times [0, L_y] \times [0, H]$, where $L_x/H = 6$, $L_y/H = 3$ and $H/h = 12$, where $h$ represents the elevation of the actuator disk center.
The scale separation and aspect ratio of the domain is chosen based on extensive analysis provided in \citet{Sathe2024}, where the impact of the numerical domain on turbulent flow statistics is examined for canopy flows.
Diameter of the disks ($D$) is kept constant $D/h = 1$ for all the simulations.
An aerodynamic roughness length of $z_0 = 10^{-6}h$ is prescribed at the lower surface via the wall-layer model. 
With the chosen value of $z_0$, the SGS pressure drag is a negligible contributor to the overall momentum balance \citep{yang2016recycling}.
The flow is in fully rough aerodynamic regime with a roughness Reynolds number $Re_\tau \equiv u_\tau h / \nu = 4.5 \times 10^6$.
The domain is discretized using a uniform Cartesian grid of $N_x \times N_y \times N_z = 576 \times 288 \times 192$.
This resolution ensures 8 grid points per diameter in the spanwise direction and 16 grid points per diameter in the vertical direction, which is shown to be adequate to accurately capture first- and second-order flow statistics \citep{Wu2011}.

\begin{figure}
  \centerline{\includegraphics[width=\textwidth]{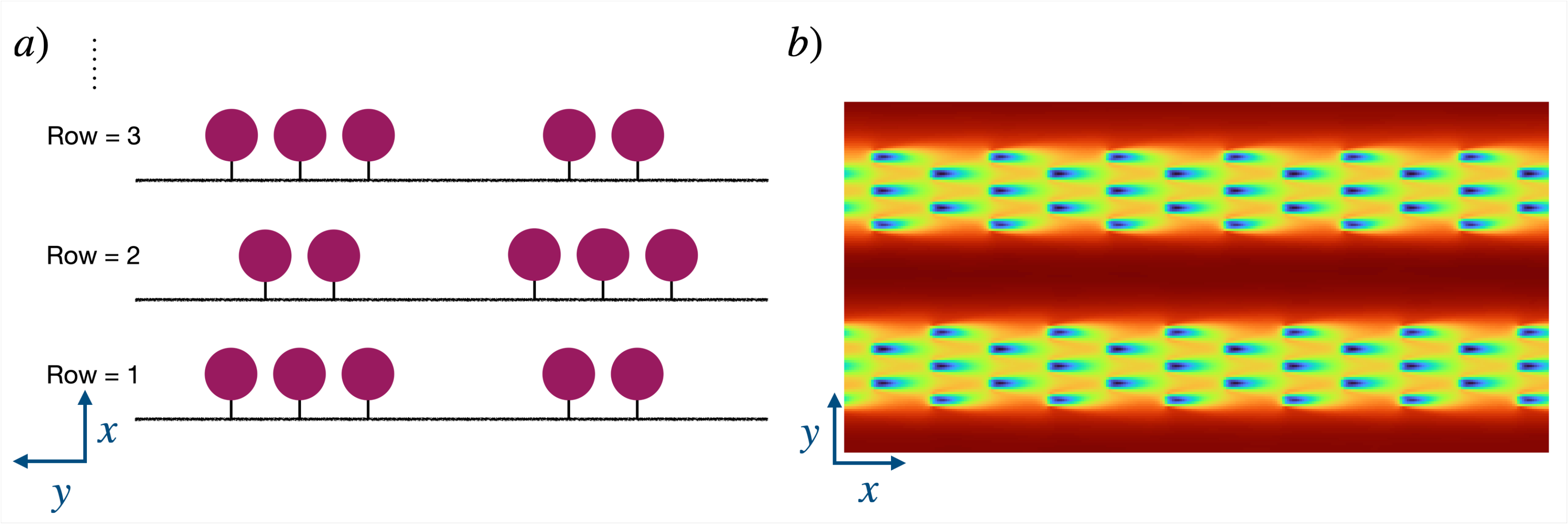}}
  \caption{a) Schematic of roughness element arrangement (not to scale; flow direction: bottom to top) and b) mean streamwise velocity at mid-element height (flow direction: left to right) for S2-5-2 case.}
\label{fig:wt_config}
\end{figure}

The cases considered in this study consist of various configurations of roughness elements which are named as detailed in equation~\ref{eq:naming scheme}.
Schematic and flow field visualization is shown for a sample case S2-5-2 in figure~\ref{fig:wt_config}.
In this naming convention, the first symbol indicates whether the elements within each column are locally staggered (denoted by `S') or aligned (denoted by `A').
The second symbol represents the spanwise gap between elements within the same column, normalized by the diameter. 
The third symbol corresponds to the total number of elements per row. 
As illustrated in figure~\ref{fig:wt_config}(a), the locally staggered configuration also involves a streamwise staggering to maintain an equal number of elements per row. 
The fourth symbol indicates the total number of wider columns, each of which may contain multiple elements in the spanwise direction.
The centers of these wider columns are uniformly spaced across the spanwise extent of the domain.
To elaborate, the case S2-5-2, which is depicted in figure~\ref{fig:wt_config}, denotes a locally staggered arrangement (S), where the spanwise gap between the centers of elements in the same wider column is twice the element diameter (S2). This configuration includes 5 elements per row (S2-5), which are distributed in two wider columns (S2-5-2).

\begin{equation}
\text { Case } \equiv \hspace{-5pt}\underbrace{(\text { Symbol 1) }}_{\begin{array}{c}
\text { \small Element } \\
\text { \small configuration }
\end{array}} \hspace{-10pt}\underbrace{(\text { Symbol 2) }}_{\begin{array}{c}
\text { \small Spanwise gap} \\
\text { \small between elements }\\
\text { \small of the same column }
\end{array}} \hspace{-12pt}- \underbrace{(\text { Symbol 3) }}_{\begin{array}{c}
\text { \small Total number } \\
\text { \small of elements }\\
\text { \small per row }
\end{array}} \hspace{-3pt}- \hspace{3pt} \underbrace{(\text { Symbol 4) }}_{\begin{array}{c}
\text { \small Number of  } \\
\text { \small larger } \\
\text { \small columns}
\end{array}} .
\label{eq:naming scheme}
\end{equation}

\section{Time averaged secondary flows and their characteristics}\label{sec: Time averaged results}
This section investigates the influence of various parameters in a multi-column roughness configuration on the polarity of secondary flows. 
Initially, four cases are examined in \S\ref{sec: Critical parameter identification}, where two cases exhibit a reversal in polarity, while the other two represent an intermediate state. 
Based on an analysis of mean kinetic energy pathways, a hypothesis is formulated which identifies the critical parameter governing polarity.
In subsequent subsections, other parameters are systematically ruled out as critical, thereby reinforcing the proposed hypothesis.

The operation of time-averaging is denoted by $\overline{(\cdot)}$, while the angled brackets $\langle \cdot \rangle$ denote horizontal averaging operation, with the subscripts indicating the direction of averaging.
\R{Deviations from the time-averaged quantities are denoted by $(\cdot)^\prime$, and the deviations from both the space and time-averaged quantities are denoted by $(\cdot)^{\prime \prime}$ \citep{Raupach1982,schmid2019volume}.}
It is important to note that all second-order statistics discussed are computed using the resolved portion of the flow field, as the contributions from the subgrid-scale components are observed to be negligible.

\subsection{Identifying the critical parameter governing secondary flow polarity} \label{sec: Critical parameter identification} 
\begin{figure}
  \centerline{\includegraphics[width=0.6\textwidth]{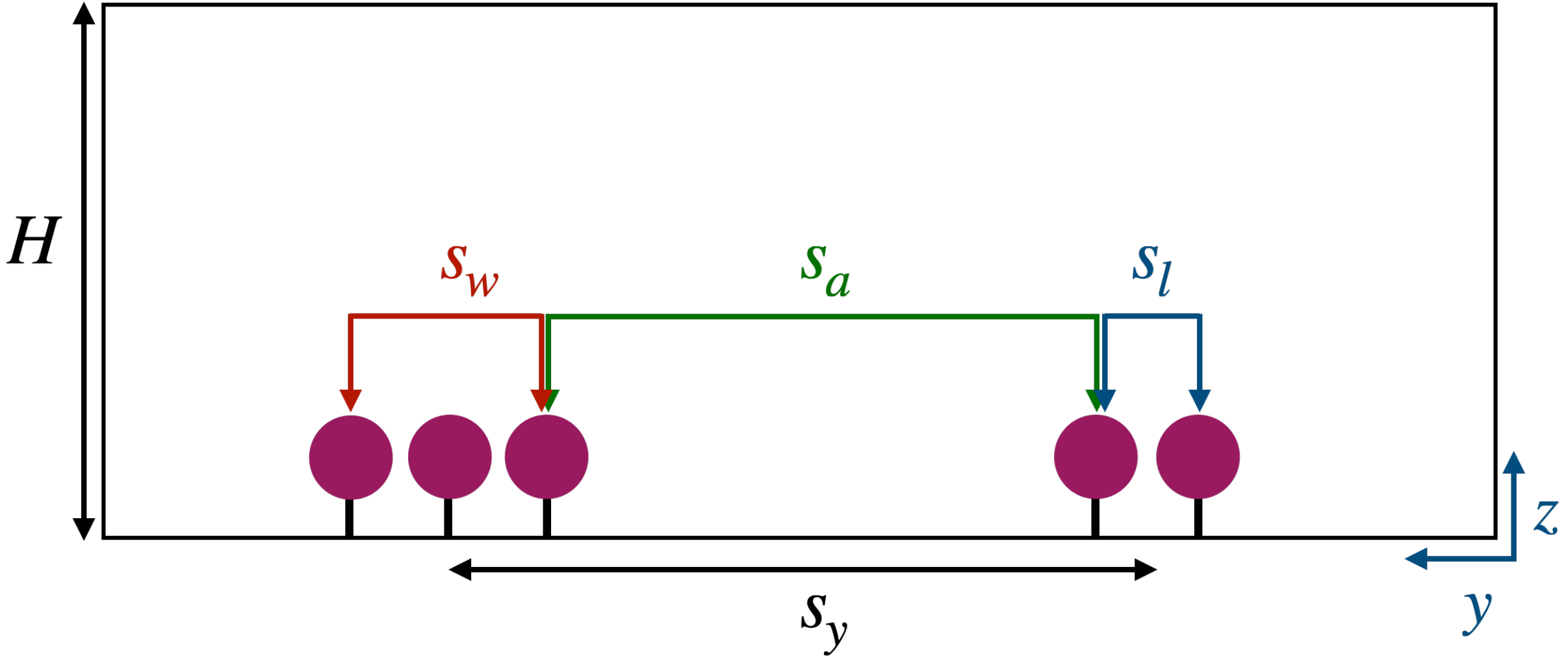}}
  \caption{Schematic of length scales considered in the roughness element arrangement. $s_l$: spanwise gap between adjacent elements of the same wider column, $s_w$: width of the wider column, $s_a$: spanwise gap between adjacent elements of different wider columns, $s_y$: spanwise gap between centers of different wider columns.}
\label{fig:length scales}
\end{figure}

In this subsection, we aim to study parameters in the multi-column roughness configuration that govern the polarity of secondary flows.
The roughness element arrangement considered involves multiple length scales of interest, as shown in figure~\ref{fig:length scales}.
Here, the length scale $s_l$ refers to the local spanwise gap between adjacent elements of the same wider column, while $s_a$ refers to the spanwise gap between adjacent elements of the neighboring wider columns. 
Since the elements are staggered in the streamwise direction as shown in figure~\ref{fig:wt_config}(a), $s_a$ remains constant across all rows.
The length scale $s_w$ refers to the total width of a column, defined as the maximum spanwise gap between elements within the same column.
Although the left column appears wider than the right in figure~\ref{fig:length scales}, $s_w$ is identical for both columns due to staggered arrangement, which can be appreciated from the mean flow field visualized in figure~\ref{fig:wt_config}(b).
The length scale $s_y$ refers to the spanwise distance between the centers of the wider columns, while $H$ is the height of the half-channel.
In this paper, the lateral length scales confined to a single wider column ($s_l$ and $s_w$) are normalized with $D$, while the length scales corresponding to different wider columns ($s_a$ and $s_y$) are normalized with $H$.
The streamwise spacing between element rows, denoted by $s_x$, is normalized by $h$.

Based on variations in these length scales, four different arrangements of roughness elements are obtained which are outlined in table~\ref{tab:staggered reversal}.
The case names reflect changes in only the second symbol, which corresponds to the length scale $s_l/D$.
As shown in the table, altering $s_l/D$ also affects parameters $s_w/D$ and $s_a/H$, due to their inherent interdependence. 
Since $s_y/H$ and $s_x/h$ have previously been identified as critical parameters influencing the polarity of secondary flows \citep{Yang2018_updated, Joshi2022}, they are fixed at a constant value across all four cases to eliminate their influence from the current analysis.

\begin{table}
  \centering
  \ra{1.3}
  \begin{tabular}{@{}llrlrlrlrlrlclclc} 
  Case & \phantom{abc} &  \multicolumn{1}{r}{$s_l/D$}& \phantom{a}   & \multicolumn{1}{c}{$s_w/D$}& \phantom{a} & \multicolumn{1}{r}{$s_a/H$}& \phantom{a}  & \multicolumn{1}{r}{$s_y/H$}& \phantom{a}  & \multicolumn{1}{r}{$s_x/h$} & \phantom{abc} & \multicolumn{1}{c}{\R{$z_0/h$}} & \phantom{a} &  \multicolumn{1}{c}{\R{$d/h$}} & \phantom{a} & \multicolumn{1}{c}{\R{$k_s/h$}}\\ \cmidrule{3-11} \cmidrule{13-17}
  S2-5-2 && 2&& 4&& 1.25&& 1.5&& 4.5&& \R{0.016} && \R{0.163} && \R{0.406}\\
  S2.75-5-2 && 2.75&& 5.5&& 1.156&&1.5&& 4.5&& \R{0.017} && \R{0.201} && \R{0.447}\\
  S3.5-5-2 && 3.5&& 7&& 1.063&&1.5&& 4.5&& \R{0.023} && \R{0.060} && \R{0.594}\\
  S4-5-2 && 4&& 8&& 1&&1.5&& 4.5&& \R{0.025} && \R{0.029} && \R{0.651}\\
  \end{tabular}
  \caption{Suite of simulations for secondary flow reversal in \S\ref{sec: Critical parameter identification}. The naming scheme for cases is defined in equation~\ref{eq:naming scheme} and length scales ($s_l$, $s_w$, $s_a$ and $s_y$) are shown in figure~\ref{fig:length scales}. $s_x$ is the streamwise gap between the element rows.}
  \label{tab:staggered reversal}
  \end{table}

\begin{figure}
  \centerline{\includegraphics[width=\textwidth]{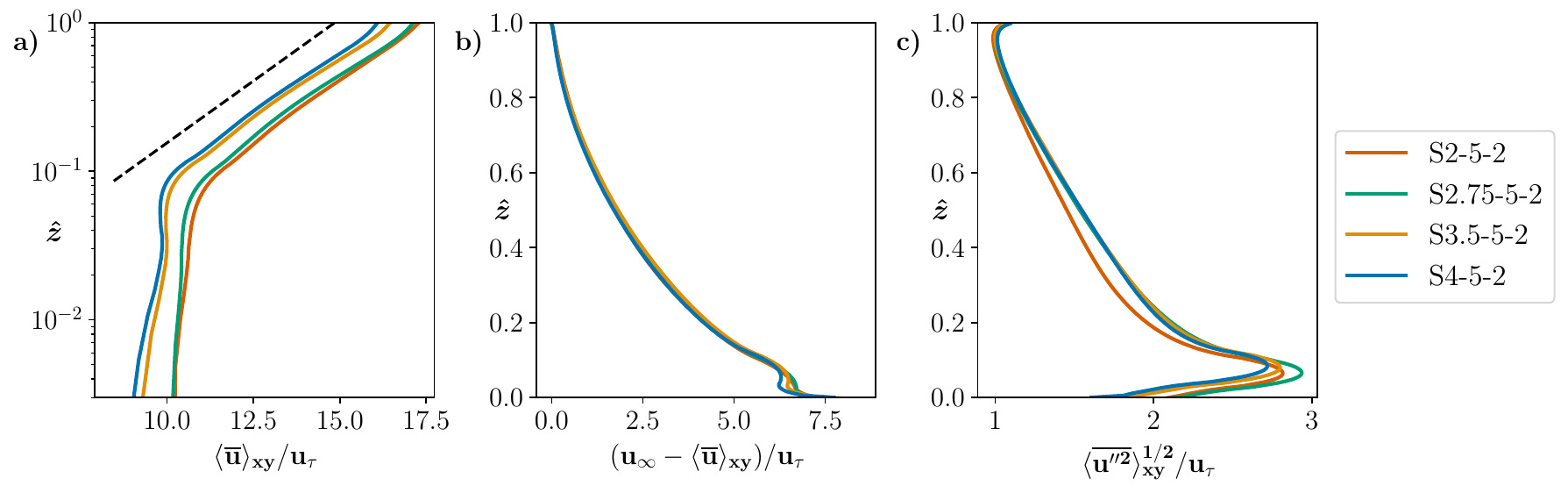}}
  \caption{\R{a) Spatially averaged mean streamwise velocity, b) velocity defect and c) root mean squared velocity profiles for the cases considered in table~\ref{tab:staggered reversal}. The black dashed line in (a) denotes the log-law slope. $\hat{z} = (z-d)/(H-d)$.}}
\label{fig:oneD profiles}
\end{figure}

\R{Figure~\ref{fig:oneD profiles} presents the horizontally averaged first- and second-order statistics for the four configurations listed in table~\ref{tab:staggered reversal}.
In rough-wall flows, the mean streamwise velocity profile in the logarithmic region is often characterized using the aerodynamic roughness length $z_0$ and the displacement height $d$, as
\begin{equation}\label{eq: log-law}
    \frac{\langle \overline{u} \rangle}{u_\tau} = \frac{1}{\kappa} ln\left(\frac{z-d}{z_0}\right) \,
\end{equation}
\noindent where $\kappa = 0.384$ is the von K\'arm\'an constant used in this study.
Owing to the pronounced spanwise heterogeneities introduced by secondary flows, as will be demonstrated later through flow visualizations, prior studies have shown that horizontally averaged profiles in such flows generally deviate from the classical log-law scaling \citep{Castro2021, Sathe2024}. 
Surprisingly, however, all four cases examined here exhibit a discernible logarithmic region, suggesting that the porous and elevated nature of the roughness elements helps preserve near-logarithmic scaling in the mean profile. 
The corresponding aerodynamic roughness lengths $z_0$ and displacement heights $d$ are listed in table~\ref{tab:staggered reversal}.
For reference, the equivalent sandgrain roughness height, computed as $k_s = z_0 e^{8.5\kappa}$, is also reported in the same table \citep{Womack2022}.
Furthermore, while the normalized velocity defect profiles collapse across all four cases, the total streamwise fluctuation profiles, comprising both temporal and dispersive contributions, exhibit a noticeable departure for case S2-5-2.}

\begin{figure}
  \centerline{\includegraphics[width=\textwidth]{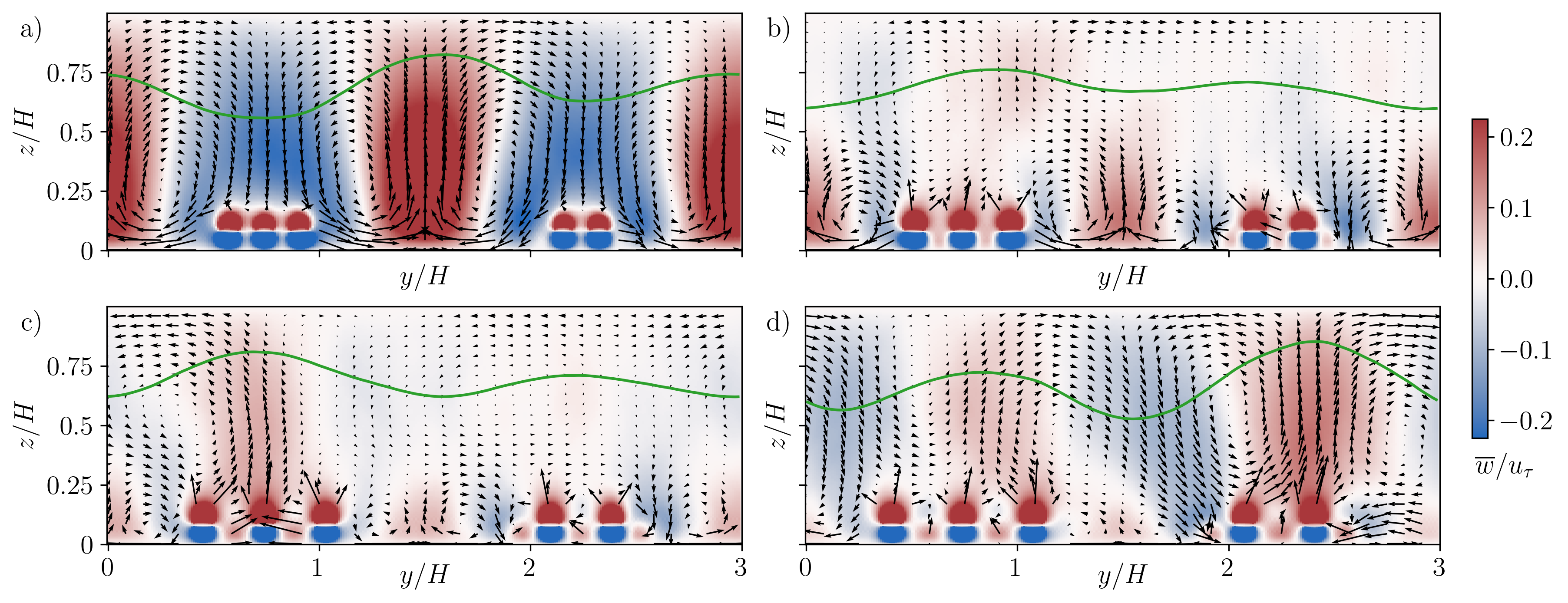}}
  \caption{Pseudocolor plot of vertical velocity for different cases mentioned in table~\ref{tab:staggered reversal}, taken at a streamwise location coinciding with the elements. a) S2-5-2, b) S2.75-5-2, c) S3.5-5-2, d) S4-5-2. Black arrows indicate the vectors of spanwise and vertical velocity. Green line indicates contour of 95\% of horizontally-averaged maximum mean streamwise velocity.}
\label{fig:mean_vort_2-3clubbed}
\end{figure}

Figure~\ref{fig:mean_vort_2-3clubbed} shows the pseudocolor plot of mean vertical velocity for these cases.
The mean flow field is averaged over repeating patterns of roughness elements in the streamwise direction to improve symmetry.
Since the structure of the mean secondary flows remains largely unchanged as the flow evolves within each repeating unit, we present only a representative $Y$–$Z$ slice taken at the streamwise location coinciding with the element rows.
Here, the obstacles appear as distinct elements with their own local updrafts and downdrafts divided at the middle of the rotor disk. 
This is due to the non-rotational nature of the actuator disk model used in this study.

It can be clearly seen that as the local spanwise gap between the elements ($s_l/D$) is varied systematically, the response of the flow due to the roughness arrangement changes significantly.
For the case S2-5-2, where elements are placed close to each other, half-channel height-scale secondary flows (here onwards referred to as delta-scale secondary flows) are distinctly seen.
Such delta-scale secondary flows are also observed in numerical studies of flow over modeled roughness \citep{Willingham2014, Anderson2015Secflow, Chung2018}, resolved roughness of different types \citep{Yang2018_updated, Hwang2018, Stroh2020} and in experimental studies \citep{Vanderwel2015, Kevin2017}, when $s_y/H \gtrsim 1.5$.
However, different polarity of these delta-scale secondary flows is also reported in these studies.
In figure~\ref{fig:mean_vort_2-3clubbed}(a), clear downdrafts are seen over the elements for case S2-5-2.
Consequently, strong updrafts are observed in the valley between the elements due to the secondary flows.
As $s_l/D$ is systematically increased, the flow response for cases S2.75-5-2 and S3.5-5-2 differs significantly.
The delta-scale vortices disappear in the mean flow for these configurations, and as a consequence, no prominent updrafts or downdrafts are observed coinciding with both the element columns symmetrically.
As $s_l/D$ is further increased, the delta-scale secondary flows reappear, but this time, strong updrafts are seen over the elements for the case S4-5-2, accompanied by strong downdrafts in the valley.
Observations of the vertical velocity at the domain center show that the extent of the updraft in the valley, originating at the wall, decreases monotonically with increasing $s_l/D$.
Figure~\ref{fig:mean_vort_2-3clubbed} also includes the 95\% contour of the horizontally averaged maximum mean streamwise velocity, highlighting the spanwise variations in the mean streamwise flow.
For the case S2-5-2, the contour shows significant modulation of mean streamwise velocity in a way such that the higher momentum fluid flows over the elements and the flow slows down significantly in the valley between the elements.
Significant reduction in the modulation of mean streamwise velocity is observed for S2.75-5-2, whereas a reversed modulation starts appearing for S3.5-5-2, where the flow slows down over the elements and speeds up in the valley. 
This reversed modulation is strengthened for the case S4-5-2, where the LMPs align with the updrafts over the elements.
This clearly showcases that as the parameter $s_l/D$ is changed, the mean secondary rolls rearrange themselves to change the locations of LMPs and HMPs with respect to the roughness arrangement.

\begin{figure}
  \centerline{\includegraphics[width=\textwidth]{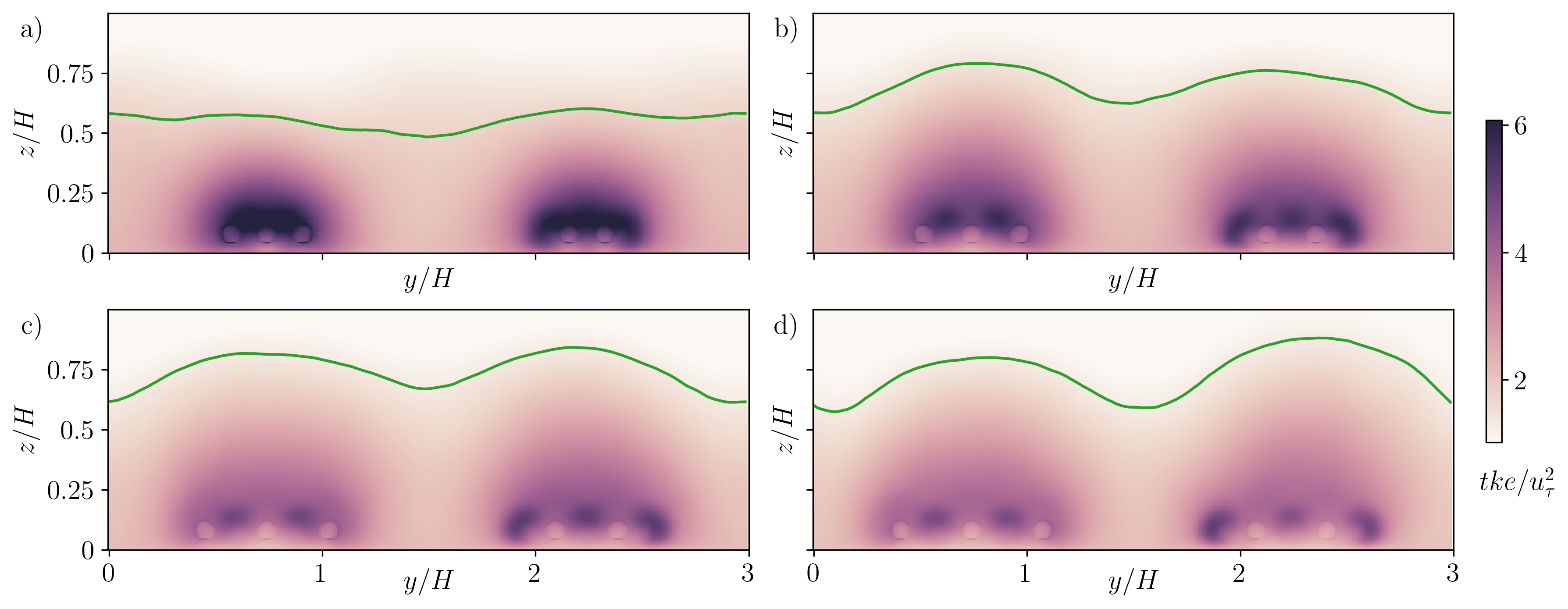}}
  \caption{Pseudocolor plot of turbulent kinetic energy (\textit{tke}) for different cases mentioned in table~\ref{tab:staggered reversal}, taken at a streamwise location coinciding with the elements. a) S2-5-2, b) S2.75-5-2, c) S3.5-5-2, d) S4-5-2. Green line indicates contour of 20\% of maximum \textit{tke} on the visualized plane.}
\label{fig:tke}
\end{figure}

Figure~\ref{fig:tke} shows the \textit{tke} at Y-Z plane coinciding with the element row for the cases discussed above.
For all the cases, maximum \textit{tke} lies in the vicinity of the roughness compared to the valley, regardless of the polarity of secondary flows.
However, the way this \textit{tke}, generated near the roughness elements, is subsequently advected throughout the flow depends strongly on the polarity of the secondary circulation.
For the case S2-5-2, due to strong downdrafts occurring over the element rows, the \textit{tke} generated by the elements is advected down and towards the valley.
This can be appreciated by looking at the contour of 20\% of maximum \textit{tke}.
Despite having two large sources in the spanwise direction that generate \textit{tke}, the 20\% contour does not show any significant hills or valleys for this case, indicating that the \textit{tke} is well-mixed in the spanwise direction.
As $s_l/D$ is increased, hills in the contour of \textit{tke} start appearing coinciding with the element rows.
This is expected, as the secondary rolls rearrange themselves to favour updrafts over the elements, which results in the generated \textit{tke} by elements being advected upwards.
This creates significant spanwise heterogeneity in the magnitude of \textit{tke} which is sustained deep in the outer layer.

While a reversal in secondary flow polarity is observed as $s_l/D$ is changed, it does not necessarily imply that $s_l/D$ is the critical parameter governing the polarity of secondary flows. 
From table~\ref{tab:staggered reversal}, it is seen that changing $s_l/D$ also changes parameters $s_w/D$ and $s_a/H$ due to their interdependence, and each of these parameters could be critical for the flow response observed in figure~\ref{fig:mean_vort_2-3clubbed}.
Furthermore, these observations are made for a locally staggered arrangement, which was chosen so that the parameter $s_l/D$ can be increased without the formation of local valleys within the same column.
However, in a staggered arrangement, as in contrast with an aligned arrangement, shear emanating from the upstream element may freely impinge on the downstream element, which can have a critical impact on the value of production term in the \textit{tke} budget equation. 
As the imbalance between production, dissipation and transport terms in the \textit{tke} budget is shown to drive the secondary flow polarity \citep{Hinze1967, Anderson2015Secflow, Hwang2018, Joshi2022}, it is unclear, thus far, whether the flow response observed is specific to the local staggered arrangement used in the study. 
Hence, further investigation is needed to accurately isolate these parameters and determine their effect on the secondary flow polarity. 
The subsequent analysis in this subsection only focuses on the cases S2-5-2 and S4-5-4 as they show clear reversal in secondary flow polarity.
While the long-time-averaged flow fields for cases S2.75-5-2 and S3.5-5-2 may give the impression that delta-scale secondary flows are absent, this is not an accurate reflection of the instantaneous flow dynamics. 
In these cases, no single pathway remains dominant over time; instead, frequent switching between pathways leads to a weakening of the secondary flow signatures in the mean visualization. 
A detailed discussion of this behavior is provided in \S\ref{sec: Instantaneous results}.

To investigate the parameter critical to the reversal in secondary flow polarity, we use the energy transport tubes introduced in \citet{Meyers2013} to identify pathways taken by the kinetic energy of the mean flow (\textit{mke}) which is extracted by a given roughness element.
Similar to the concept of classical mass transport tubes, energy transport tubes are defined such that no fluxes of energy exist over the tube mantle. 
The flux vector of \textit{mke} can be defined based on its transport equation,

\begin{equation} \label{eq: enery-transport}
\frac{\partial}{\partial x_j}\left(\overline{E}_{j}\right)=-\frac{1}{\rho}\frac{\partial \overline{u}_i \overline{p}_\infty}{\partial x_i}+ \overline{u_i^{\prime} u_j^{\prime}} \frac{\partial \overline{u}_i}{\partial x_j}-2 \nu \overline{S}_{i j} \overline{S}_{i j}+\frac{1}{\rho}\overline{u}_i \overline{F}_i \ ,
\end{equation}

\noindent where

\begin{equation} \label{eq: energy flux vector}
\overline{E}_{j}= \overline{K} \ \overline{u}_j+\left(\frac{\overline{p}}{\rho} \delta_{ij} + \overline{u_i^{\prime} u_j^{\prime}}-2 v \overline{S}_{i j}\right) \overline{u}_i,
\end{equation}

\noindent is the mean kinetic energy flux vector per unit mass, $\overline{K} = \left( \overline{u}_i \overline{u}_i / 2 \right)$ is the \textit{mke}, $\overline{S}_{ij}$ is the mean strain rate tensor and $\overline{p}$ represents the part of pressure responsible for pressure gradient other than the externally applied pressure gradient ($\overline{p}_\infty)$.
Equation~\ref{eq: energy flux vector} shows that the energy transport tube differs from classical mass transport tube due to the pressure, turbulence and viscous transport terms.
As the Reynolds number for all the cases under investigation is very high, the contribution of viscosity is neglected.
From equation~\ref{eq: enery-transport}, it is evident that the right-hand side contains both sources and sinks of energy. 
Therefore, unlike mass transport tubes, the total energy within an energy transport tube may not be conserved \citep{Meyers2013}.

\begin{table}
  \centering
  \ra{1.3}
  \begin{tabular}{@{}llrlrlrlrlr} 
  Case & \phantom{abc} &  \multicolumn{1}{r}{$s_l/D$}& \phantom{ab}   & \multicolumn{1}{c}{$s_w/D$}& \phantom{ab} & \multicolumn{1}{r}{$s_a/H$}& \phantom{ab}  & \multicolumn{1}{r}{$s_y/H$}& \phantom{ab}  & \multicolumn{1}{r}{$s_x/h$} \\ \cmidrule{3-11}
  A0-3-3 && N/A && N/A&& 1  &&   1&& 4.5\\
  A0-2-2 && N/A && N/A&& 1.5&& 1.5&& 4.5\\
  \end{tabular}
  \caption{Additional suite of simulations for energy tube evolution. The naming scheme is defined in equation~\ref{eq:naming scheme}. The length scales are same as table~\ref{tab:staggered reversal}.}
  \label{tab:Etube-classics}
  \end{table}

Before analyzing the mean energy transport tubes for cases S2-5-2 and S4-5-2, we first consider simpler aligned cases to understand the preferential pathways for \textit{mke} entrainment for the range of $s_a/H$ and $s_y/H$ values under investigation, which are detailed in table~\ref{tab:staggered reversal}. 
\citet[see fig. 10]{Meyers2013} showcased variations in the shapes of energy tubes with varying $s_y/D$.
In their investigation for aligned configuration, they considered different values of $s_y/D$, ranging from 4.54 to 10.5.
They showcased that as $s_y/D$ is progressively increased, the elements favor spanwise entrainment of \textit{mke} compared to the vertical entrainment.
When normalized by the half-channel height, i.e. considering $s_y/H$ instead of $s_y/D$, this parameter range corresponds to 0.454 to 1.05 for their study. 
However, it is well known that as $s_y/H$ is increased further, delta-scale secondary flows appear and strong inhomogeneities in Reynolds stresses are observed associated with these secondary motions \citep{Willingham2014, Vanderwel2015, Yang2018_updated}.
Thus, considering the contribution of advection and Reynolds stresses in equation~\ref{eq: energy flux vector}, it is unclear whether the preferential spanwise entrainment of \textit{mke} with increasing $s_y/H$ continues for $s_y/H > 1$.
To investigate this, we run two cases with $s_y/H = 1 \ \text{and} \ 1.5$, as detailed in table~\ref{tab:Etube-classics}.
These cases have aligned configuration with only 1 element per column in the spanwise direction to mimic the cases considered in \citet{Meyers2013}.
The streamwise gap between the elements is kept equal to the cases shown in table~\ref{tab:staggered reversal} to facilitate comparison with S2-5-2 and S4-5-2.

\begin{figure}
  \centerline{\includegraphics[width=\textwidth]{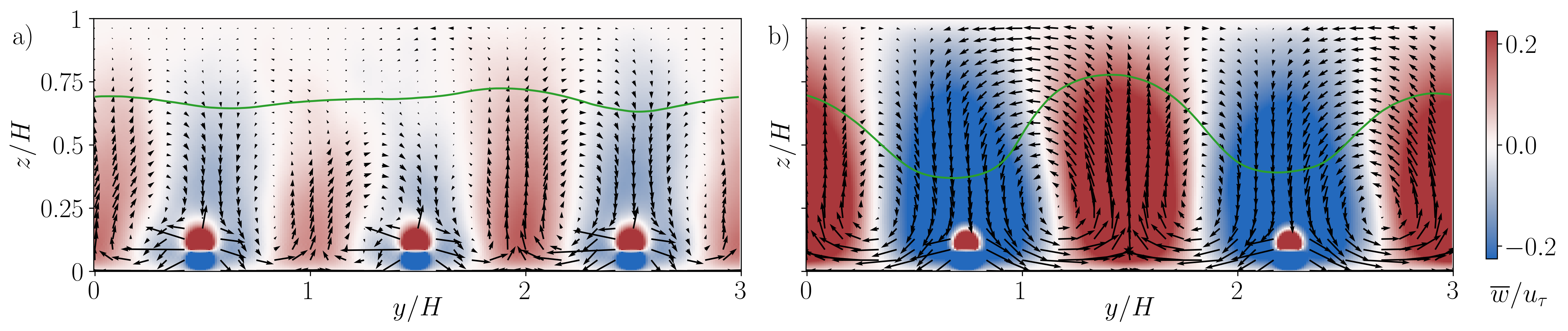}}
  \caption{Pseudocolor plot of vertical velocity for different cases mentioned in table~\ref{tab:Etube-classics}, taken at a streamwise location coinciding with the elements. a) A0-3-3, b) A0-2-2. Black arrows indicate the vectors of spanwise and vertical velocity. Green line indicates contour of 95\% of maximum mean streamwise velocity.}
\label{fig:mean_vort_classics}
\end{figure}

\begin{figure}
  \centerline{\includegraphics[width=0.7\textwidth]{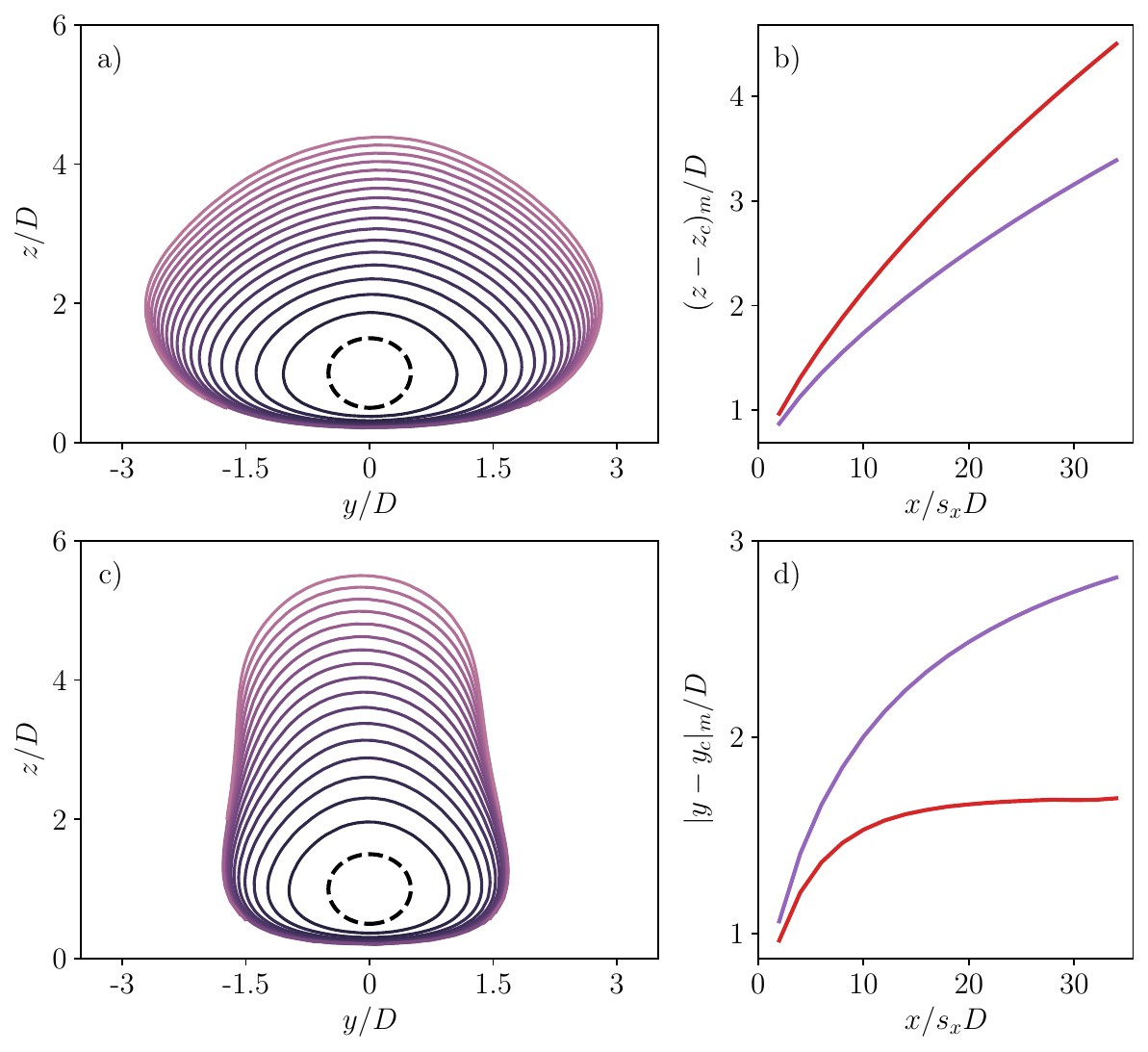}}
  \caption{Mean energy transport tubes for a) A0-3-3 and c) A0-2-2. Element location is shown in black dashed line. Tube outlines are plotted at different upstream locations $x=-ns_xD$, with $n=2, 4, ..., 34$. The outline color gradient indicates the upstream distance from the element, with lighter colors representing locations farther upstream. b) Maximum height and d) maximum spanwise distance at which the \textit{mke} is entrained by the element relative to the element center. Purple: A0-3-3, red: A0-2-2.}
\label{fig:Etubes-classics}
\end{figure}

Figure~\ref{fig:mean_vort_classics} presents the pseudocolor plot of vertical velocity and modulation of mean streamwise velocity for these cases.
For case A0-3-3, which corresponds to $s_y/H=1$, weak downdrafts are observed over the elements and updrafts are observed in the valley.
However, the effect of these circulations on the mean streamwise velocity is minimal.
On the other hand, for the case A0-2-2, which corresponds to $s_y/H=1.5$, strong delta-scale secondary flows are observed with HMPs and corresponding downdrafts coinciding with the elements.
The effect of these circulations on the mean energy transport tubes is depicted in figure~\ref{fig:Etubes-classics}.
For the case A0-3-3, the energy tube expands in both the vertical and the spanwise direction equally.
We note that this shape differs from the case with $s_y/H=1.05$ considered in \citet{Meyers2013} due to the comparatively smaller streamwise gap considered in our study.
For A0-3-3, initially, the growth rate is higher in the spanwise direction.
However, as the upstream distance increases, the lower half of the energy tube starts converging to a single line indicating diminishing spanwise growth, whereas the vertical growth is still noticeable till $x=-34s_xD$.
However, in contrast, the energy tube for A0-2-2 is noticeably narrower and taller.
By analyzing the maximum vertical and spanwise distance of the location of the \textit{mke} which is entrained by the elements, which is shown in figure~\ref{fig:Etubes-classics}(b, d), it is clear that the case A0-2-2 favors vertical entrainment of \textit{mke} over the spanwise entrainment compared to A0-3-3.
This showcases that the observed trend of increasing spanwise entrainment with increasing $s_y/H$ for $s_y/H \lesssim 1$ is reversed due to secondary circulations as $s_y/H$ increases beyond 1.
This is expected, as in this case, the secondary circulations align the higher-momentum fluid with the region above the elements, and strong downdrafts facilitate its downward transport.
From this analysis, we can see that the shape and evolution of the mean energy transport tube encodes the information of polarity of secondary circulations.
This is because as we move away from the elements, the contribution of pressure transport and turbulent transport in equation~\ref{eq: energy flux vector} decreases.
Thus, the location of extremities of the tube far away from the elements indicates the direction of advection.
Specifically, a narrower and taller tube indicates downdraft over the elements, which forms a circulation such that the flow is advected from the elements towards the valley (see, e.g., figure~\ref{fig:mean_vort_2-3clubbed}(a)).
In contrast, a shorter and wider tube indicates spanwise advection of \textit{mke}, which forms a circulation such that the flow is advected from the valley towards the elements, which results in updrafts over the elements (see, e.g., figure~\ref{fig:mean_vort_2-3clubbed}(d)).
For the cases such as A0-3-3 where the energy tube is equally wider and taller, formation of HMPs and LMPs is not observed which is reflected in weak modulation of mean streamwise velocity shown in figure~\ref{fig:mean_vort_classics}(a).
It should also be noted that, for both A0-3-3 and A0-2-2, the two parameters $s_a/H$ and $s_y/H$ are identical. 
Thus, this change in entrainment pathway of \textit{mke} with changing $s_y/H$ can also be interpreted as a change due to changing $s_a/H$, which denotes the gap between adjacent elements in the spanwise direction.

\begin{figure}
  \centerline{\includegraphics[width=0.7\textwidth]{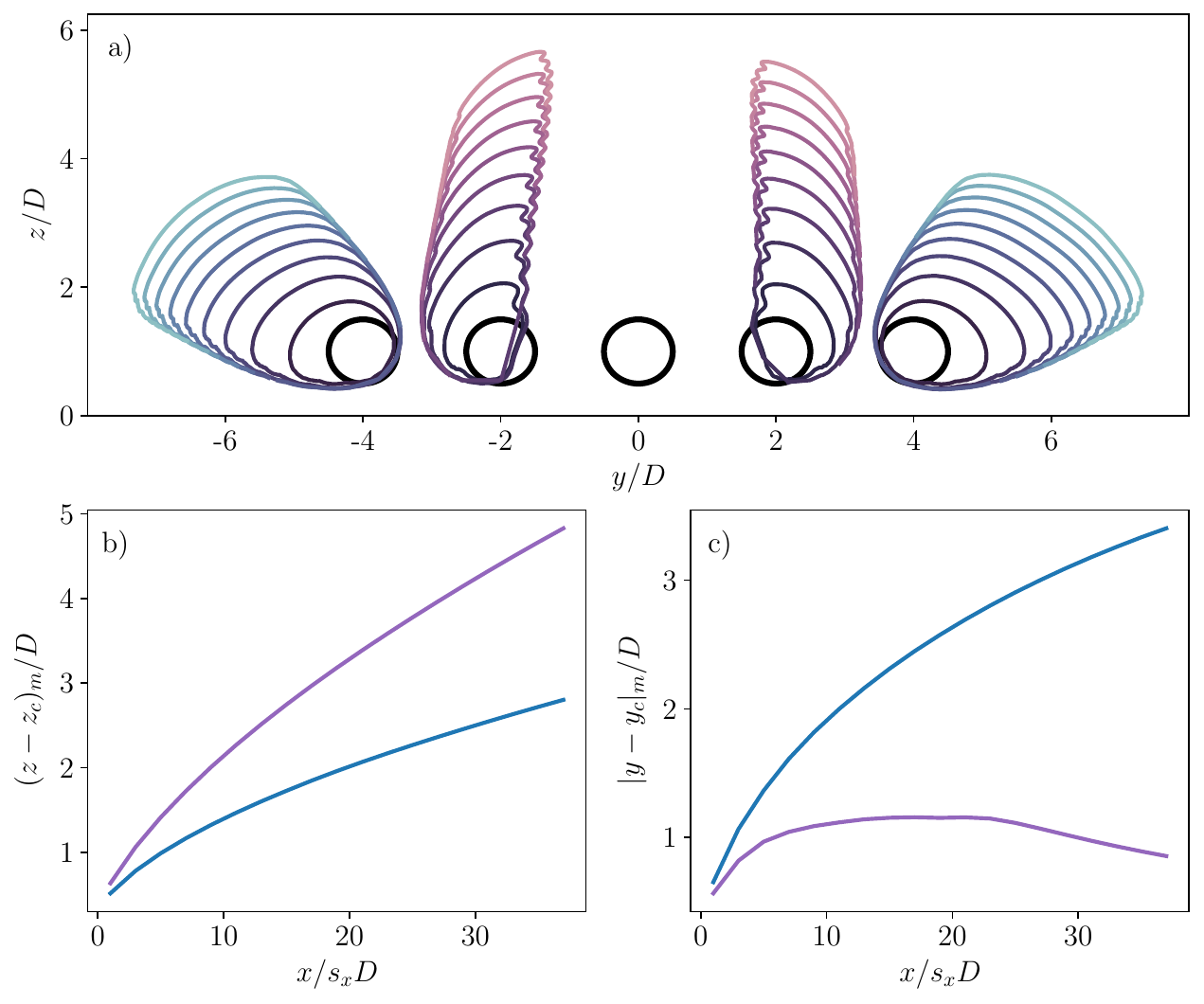}}
  \caption{a) Mean energy transport tubes for S2-5-2 (purple) and S4-5-2(blue). Elements are shown in black lines with locations for S2-5-2 and S4-5-2 being $y/D=\{-2, 0, 2\}$ and $y/D=\{-4, 0, 4\}$ respectively. Tube outlines are plotted at different upstream locations $x=-ns_xD$, with $n=3, 5, ..., 37$. The outline color gradient indicates the upstream distance from the element, with lighter colors representing locations farther upstream. b) Maximum height and d) maximum spanwise distance at which the \textit{mke} is entrained by the element relative to the element center. Purple: S2-5-2, blue: S4-5-2}
\label{fig:Etubes-secondary}
\end{figure}

Figure~\ref{fig:Etubes-secondary} shows the mean energy transport tubes for cases S2-5-2 and S4-5-2 along with the maximum vertical and spanwise distance at which the \textit{mke} is entrained by the elements.
To enable direct comparison, the two layouts are overlaid, and we focus on the column segment containing three elements, where the central elements coincide and the edge elements are offset according to their respective $s_l/D$ values.
To maintain visual clarity, the transport tubes associated with the central element are omitted, as the focus here is on the entrainment behavior of the edge elements.
The figure shows a clear contrast in the entrainment mechanisms between the two cases. 
In case S2-5-2, the elements predominantly entrain \textit{mke} from above, indicating a vertically dominant entrainment pathway. 
In contrast, case S4-5-2 exhibits stronger spanwise entrainment. 
Interestingly, this shift in entrainment direction occurs despite both configurations sharing the same $s_y/H$ value, as their central elements are aligned at $y/D = 0$. 
This indicates that a parameter other than $s_y/H$ is responsible for the different entrainment pathways observed for these cases.
As discussed in the previous paragraph, a change in $s_a/H$—which differs from $s_y/H$ in these configurations—may also influence the \textit{mke} entrainment pattern.
Specifically, we noted that the cases may fall into different $s_a/H$ regimes, wherein the elements tend to favor either vertical or lateral entrainment.
Table~\ref{tab:staggered reversal} highlights this distinction, showing that $s_a/H$ equals 1.25 for S2-5-2 and 1.00 for S4-5-2, which may place them in separate $s_a/H$ regimes.
This shift in $s_a/H$ appears to be the driving factor behind the distinct entrainment patterns observed in S2-5-2 and S4-5-2, which ultimately result in the reversal of mean secondary flow polarity seen in figure~\ref{fig:mean_vort_2-3clubbed}.
Thus, we hypothesize that $s_a/H$ is the critical parameter in a multi-column roughness configuration that governs the polarity of secondary flows.
In subsequent subsections, we will disprove $s_l/D$, staggered configuration and $s_w/D$ as being crucial to the polarity reversal observed in figure~\ref{fig:mean_vort_2-3clubbed}, thereby reinforcing our hypothesis.

\subsection{Disproving the local spanwise gap ($s_l/D$) as the critical parameter}\label{sec: disproving sl/D}

In this subsection, we provide further evidence supporting our conclusion that $s_a/H$ is the critical parameter by showing that $s_l/D$ does not play a decisive role.
At a first glance, based on the plots shown in figure~\ref{fig:mean_vort_2-3clubbed}(a, d), one might speculate that increasing $s_l/D$, which results in a significantly larger local spanwise gap between the elements in case S4-5-2 compared to S2-5-2, could influence the preferential fluid pathways locally. 
A narrower spanwise gap (S2-5-2) may require vertical entrainment of the flow, resulting in a global downdraft over the elements, which may not be necessary when a wider spanwise gap is available for the flow (S4-5-2). 
To disprove this intuition, we present four new cases along with cases S2-5-2 and S3.5-5-2 considered previously, as outlined in table~\ref{tab:disproving sl/D}.
\begin{figure}
  \centerline{\includegraphics[width=\textwidth]{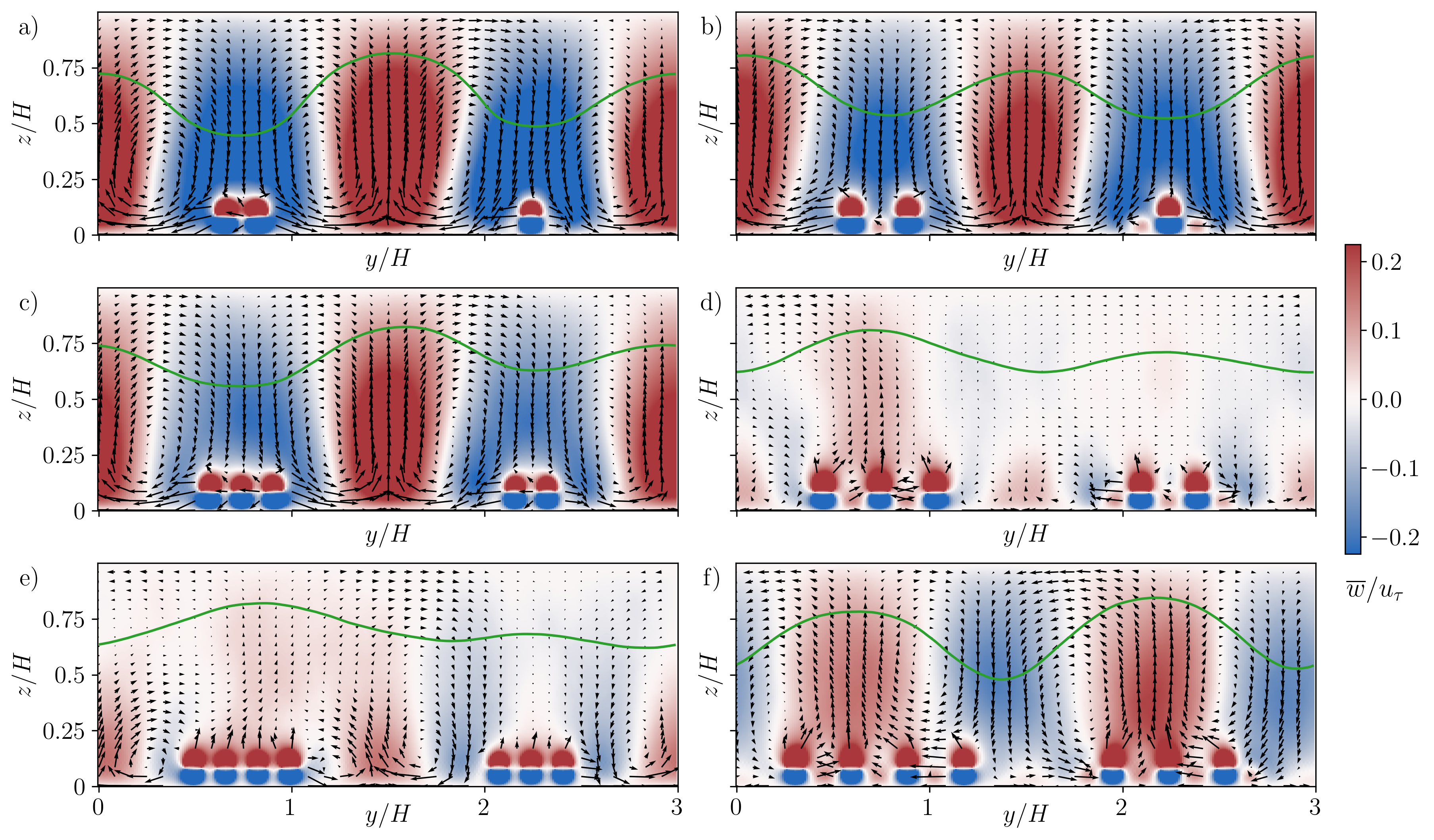}}
  \caption{Pseudocolor plot of vertical velocity for different cases mentioned in table~\ref{tab:disproving sl/D}, taken at a streamwise location coinciding with the elements. a) S2-3-2, b) S3.5-3-2, c) S2-5-2, d) S3.5-5-2, e) S2-7-2, f) S3.5-7-2. Black arrows indicate the vectors of spanwise and vertical velocity. Green line indicates contour of 95\% of maximum mean streamwise velocity.}
\label{fig:mean_vort_1-4clubbed}
\end{figure}

\begin{table}
  \centering
  \ra{1.3}
  \begin{tabular}{@{}llrlrlrlrlr} 
  Case & \phantom{abc} &  \multicolumn{1}{r}{$s_l/D$}& \phantom{ab}   & \multicolumn{1}{c}{$s_w/D$}& \phantom{ab} & \multicolumn{1}{r}{$s_a/H$}& \phantom{ab}  & \multicolumn{1}{r}{$s_y/H$}& \phantom{ab}  & \multicolumn{1}{r}{$s_x/h$} \\ \cmidrule{3-11}
  S2-3-2   && 2   && 2    && 1.417&&  1.5&& 4.5\\
  S3.5-3-2 && 3.5 && 3.5  && 1.354&&  1.5&& 4.5\\
  S2-5-2   && 2   && 4    && 1.25 &&  1.5&& 4.5\\
  S3.5-5-2 && 3.5 && 7    && 1.063&&  1.5&& 4.5\\
  S2-7-2   && 2   && 6    && 1.083&&  1.5&& 4.5\\
  S3.5-7-2 && 3.5 && 10.5 && 0.771&&  1.5&& 4.5\\
  \end{tabular}
  \caption{Suite of simulations for \S\ref{sec: disproving sl/D}. The naming scheme for cases is defined in equation~\ref{eq:naming scheme}. The length scales are same as table~\ref{tab:staggered reversal}.}
  \label{tab:disproving sl/D}
  \end{table}

Figure~\ref{fig:mean_vort_1-4clubbed} presents the pseudocolor plot of vertical velocity for the cases listed in table~\ref{tab:disproving sl/D}.
We first focus on the cases where $s_l/D = 2$.
Although these cases share the same $s_l/D$, they have different $s_a/H$ values due to the increasing number of local columns.
For cases S2-3-2 and S2-5-2, distinct downdrafts are observed above the elements.
Additionally, the modulation of mean streamwise velocity shows formation of HMPs over the elements and LMPs in the valley.
This behavior contrasts with case S2-7-2, where intermediate secondary flows occur, where the symmetry of the secondary flows breaks down.
The modulation of mean streamwise velocity is also markedly different for S2-7-2, as no distinct HMPs or LMPs coincide symmetrically with the elements.
Thus, a clear transition is observed: from HMPs over the elements in S2-3-2 and S2-5-2 to intermediate secondary flows in S2-7-2, as the value of $s_a/H$ decreases, while keeping $s_l/D$ constant.
Similarly, for cases with $s_l/D = 3.5$, we observe a reversal in flow polarity.
The flow transitions from HMPs over the elements in S3.5-3-2, to intermediate secondary flows in S3.5-5-2, and to LMPs over the elements in S3.5-7-2, as $s_a/H$ decreases.
These six cases can together be ranked in decreasing order of $s_a/H$ as $\text{S2-3-2} > \text{S3.5-3-2} > \text{S2-5-2} > \text{S3.5-5-2} \approx \text{S2-7-2} > \text{S3.5-7-2}$.
In this sequence, we observe a progressive transition from HMPs over the elements, to intermediate secondary flows, to LMPs over the elements, independent of changes in the local spanwise gap.
Thus, from the observations of different flow response for the same $s_l/D$ (cases: S3.5-3-2, S3.5-5-2, and S3.5-7-2) and similar flow response for different $s_l/D$ (cases: S2-3-2 and S3.5-3-2), we conclude that $s_l/D$ is not the critical parameter driving the reversal in secondary flow polarity observed in \S\ref{sec: Critical parameter identification}.
The variations in flow response observed with decreasing $s_a/H$ provide further reinforcement for our conclusion that $s_a/H$ is the key parameter governing this behavior.

\subsection{Disproving the local element alignment as the critical parameter}\label{sec: disproving alignment}
In this subsection, we showcase that the secondary flow polarity reversal observed in \S\ref{sec: Critical parameter identification} is not limited to the staggered configuration within a wider column.
To prove this, three new cases are simulated with aligned configuration as outlined in table~\ref{tab:disproving alignment}.
In these cases, the variation in the parameter $s_l/D$ is kept modest compared to the cases considered in \S\ref{sec: Critical parameter identification} to prevent the formations of local valleys within a wider column.
However, by including 4 elements per wider column, different regimes of $s_a/H$ are obtained despite the limited variation in $s_l/D$.
If the polarity reversal observed in \S\ref{sec: Critical parameter identification} depends solely on $s_a/H$ and is independent of the staggered configuration used, a corresponding shift in the secondary flow response is anticipated for these aligned cases as well.

\begin{figure}
  \centerline{\includegraphics[width=\textwidth]{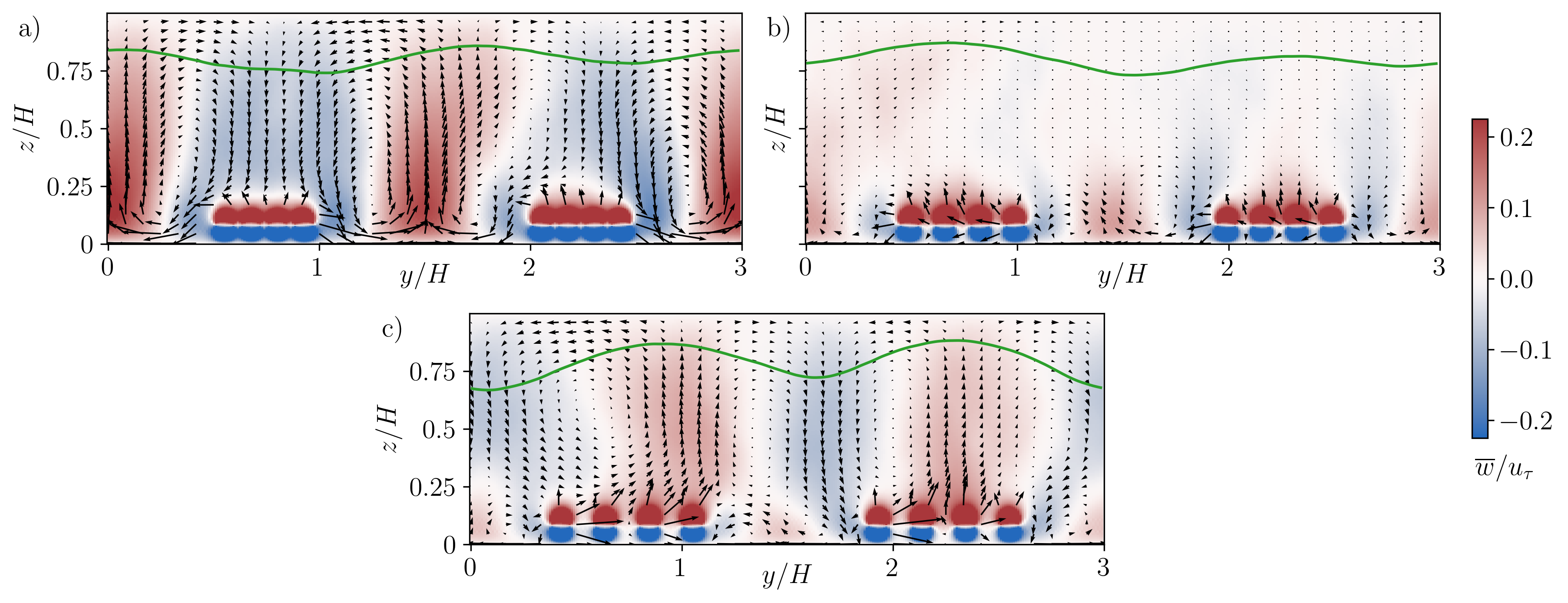}}
  \caption{Pseudocolor plot of vertical velocity for different cases mentioned in table~\ref{tab:disproving alignment}, taken at a streamwise location coinciding with the elements. a) A1.5-8-2, b) A2-8-2, c) A2.5-8-2. Black arrows indicate the vectors of spanwise and vertical velocity. Green line indicates contour of 95\% of maximum mean streamwise velocity.}
\label{fig:mean_vort_4aligned}
\end{figure}

Figure~\ref{fig:mean_vort_4aligned} shows the flow response for these three cases.
For the case A1.5-8-2, clear downdrafts are observed over the elements and updrafts are observed within the valley.
However, these circulations have only a modest impact on the modulation of mean streamwise velocity, with the formation of weak HMPs over the elements and weak LMPs in the valley.
As the ratio $s_a/H$ decreases, intermediate secondary flows are observed for case A2-8-2, where no significant updrafts or downdrafts coincide with the element locations.
With a further reduction in $s_a/H$, pronounced updrafts are observed over the elements, accompanied by the formation of LMPs over the elements and HMPs within the valley.
Although fully symmetric secondary flows are not observed in this case, the flow response between cases A1.5-8-2 and A2.5-8-2 is markedly different.
These findings indicate that as $s_a/H$ decreases, there is a transition from downdrafts to updrafts over the elements for the aligned configuration, consistent with the observations reported in \S\ref{sec: Critical parameter identification} and \S\ref{sec: disproving sl/D} for the staggered configuration.
This demonstrates that the observed flow response in \S\ref{sec: Critical parameter identification} is not limited to the local alignment of the elements within a wider column.
However, it is important to ensure that the alignment does not create significant local valleys within the wider column. 
If such valleys form, the very concept of a ``wider column" may break down, which may lead to a disruption in the expected flow behavior.

\begin{table}
  \centering
  \ra{1.3}
  \begin{tabular}{@{}llrlrlrlrlr} 
  Case & \phantom{abc} &  \multicolumn{1}{r}{$s_l/D$}& \phantom{ab}   & \multicolumn{1}{c}{$s_w/D$}& \phantom{ab} & \multicolumn{1}{r}{$s_a/H$}& \phantom{ab}  & \multicolumn{1}{r}{$s_y/H$}& \phantom{ab}  & \multicolumn{1}{r}{$s_x/h$} \\ \cmidrule{3-11}
  A1.5-8-2   && 1.5   && 4.5    && 1.125 &&  1.5 && 4.5\\
  A2-8-2     && 2     &&  6     && 1     &&  1.5 && 4.5\\
  A2.5-8-2   && 2.5   && 7.5    && 0.875 &&  1.5 && 4.5\\
  \end{tabular}
  \caption{Suite of simulations for \S\ref{sec: disproving alignment}. The naming scheme for cases is defined in equation~\ref{eq:naming scheme}. The length scales are same as table~\ref{tab:staggered reversal}.}
  \label{tab:disproving alignment}
  \end{table}

\subsection{Disproving the width of the wider column ($s_w/D$) as the critical parameter}\label{sec: disproving column-width}

In this subsection, we discuss why the secondary flow polarity reversal observed in \S\ref{sec: Critical parameter identification} is not due to the width of wider columns.
For this, we first refer the reader to the cases considered in table~\ref{tab:Etube-classics} and the corresponding energy tube visualization in figure~\ref{fig:Etubes-classics}.
For these cases, there's no change in the width of the columns, as each column contains only a single element.
Nevertheless, distinct patterns of \textit{mke} entrainment emerge when the parameter $s_a/H$ is varied.
This demonstrates that even in the absence of changes in $s_w/D$, different flow responses can be observed by varying $s_a/H$.

Next, we discuss the nature of coupling between parameters $s_l/D$ and $s_w/D$.
For cases with an equal number of elements per row, as outlined in table~\ref{tab:staggered reversal}, there exists a direct coupling between  $s_l/D$ and $s_w/D$. 
Specifically, any modification in $s_l/D$ necessitates a corresponding adjustment in $s_w/D$, precluding independent variation of these parameters.
Therefore, $s_w/D$ can be regarded as a local parameter, concerned only with a given wider column.
In \S\ref{sec: disproving sl/D}, it is demonstrated that $s_l/D$ does not play a crucial role in determining secondary flow polarity.
Given that $s_l/D$ influences only the local interactions within a wider column, this implies that, within the studied spanwise sparsity range (i.e., $s_y/H=1.5$), the elements' local spanwise columnar interactions are not crucial to the global secondary flow response.
This suggests that the different \textit{mke} entrainment patterns observed in figure~\ref{fig:Etubes-secondary} are not critically affected by these local interactions.
Since $s_w/D$ also governs local interactions, it can be inferred that the distinct \textit{mke} entrainment pathways observed in this figure are not attributable to changes in $s_w/D$.
Therefore, $s_w/D$ is not a critical parameter governing the secondary flow polarity reversal observed in \S\ref{sec: Critical parameter identification}.
We further add that the coupling between $s_l/D$ and $s_a/H$ is of different nature and is dependent through the parameter $s_y/H$.
Due to this dependence on $s_y/H$, $s_a/H$ governs interactions between neighboring wider columns, rather than local interactions within a single wider column.
Independent variations in $s_l/D$ and $s_a/H$ can be achieved by changing $s_y/H$ while maintaining the same number of elements per row.
Consequently, while it is reasonable to conclude that $s_w/D$ is not a critical parameter—given that $s_l/D$ is non-critical—the same reasoning cannot be applied to $s_a/H$.

Overall, the analysis presented in this section clearly demonstrates that variations in the parameter $s_a/H$ are the primary driver behind the reversal in secondary flow polarity observed in figure~\ref{fig:mean_vort_2-3clubbed}.
This finding highlights that, in multi-column configurations, roughness elements located at the edges of the columns can play a decisive role in setting the polarity of secondary flows. 
While the polarity reversal is evident in the long-time-averaged flow field, the extent to which these structures persist in time remains an open question. 
In particular, it is unclear whether the presence (figure~\ref{fig:mean_vort_2-3clubbed}(a,d)) or absence (figure~\ref{fig:mean_vort_2-3clubbed}(b,c)) of counter-rotating delta-scale secondary flows in the mean reflects fundamentally different instantaneous dynamics. 
We address this aspect in the following section.

\section{Instantaneous secondary flows and their characteristics}\label{sec: Instantaneous results}
In the previous section, we examined the long-time-averaged secondary flows and the reversal of their polarity. 
However, certain configurations, such as those shown in figure~\ref{fig:mean_vort_2-3clubbed}(b, c), exhibited irregular or seemingly disrupted secondary flows despite prolonged averaging.
Since domain-scale secondary flows are artifacts of time-averaging that obscure the underlying transient dynamics \citep{Kevin2017}, a closer examination of the unsteady behavior is needed to better understand these observations.
\citet{Anderson2019} demonstrated that secondary flows can undergo instantaneous reversals driven by chaotic and non-periodic dynamics.
Similarly, \citet{Vanderwel2019} found that secondary flows emerge from the non-homogeneous distribution of smaller vortices, which meander in location but collectively form large-scale secondary motions in long-time-averaged visualizations.
These findings raise the question of whether different roughness arrangements in our study influence the organization of smaller vortices, their meandering behavior, and the propensity for non-periodic reversals in secondary flows.
Understanding these tendencies could provide insight into why large-scale, time-averaged secondary flows are observed in some cases but not in others.
Therefore, in this section, we analyze the instantaneous structure of secondary flows using flow field visualization, conditional averaging, and probability density functions to better understand their transient behavior.
\R{Here, we refer to the instantaneous counterparts of LMPs and HMPs as low-momentum regions (LMRs) and high-momentum regions (HMRs), respectively.}

\R{In the analysis that follows, $\widehat{(\cdot)}$ denotes low-pass filtered quantities, and all time scales are normalized by $h/u_\tau$, represented using the superscript $(\cdot)^*$.
The conditionally averaged fields are denoted by $\overline{(\cdot)}_c$.}

\subsection{Visualization of instantaneous flow field}\label{sec: Inst visualization}
In this subsection, we examine the instantaneous flow field behavior of case S3.5-5-2.
Figure~\ref{fig:mean_vort_2-3clubbed}(c) presents the long-time-averaged flow field for this case, and the prominent observations are described briefly here.
Unlike the cases in figure~\ref{fig:mean_vort_2-3clubbed}(a, d), which exhibit symmetrical secondary flows, this case displays a pronounced asymmetry, characterized by a strong updraft in the left column and a significantly weaker updraft in the right column. 
Consequently, symmetrical delta-scale counter-rotating vortices are absent. 
The time-averaged field also reveals that an updraft remains confined near the wall at the domain center, with a weak downdraft positioned above it. 
Additionally, the mean streamwise velocity contour indicates that an HMP aligns with the domain center, while an LMP is positioned over the roughness elements.

\begin{figure}
  \centerline{\includegraphics[width=\textwidth]{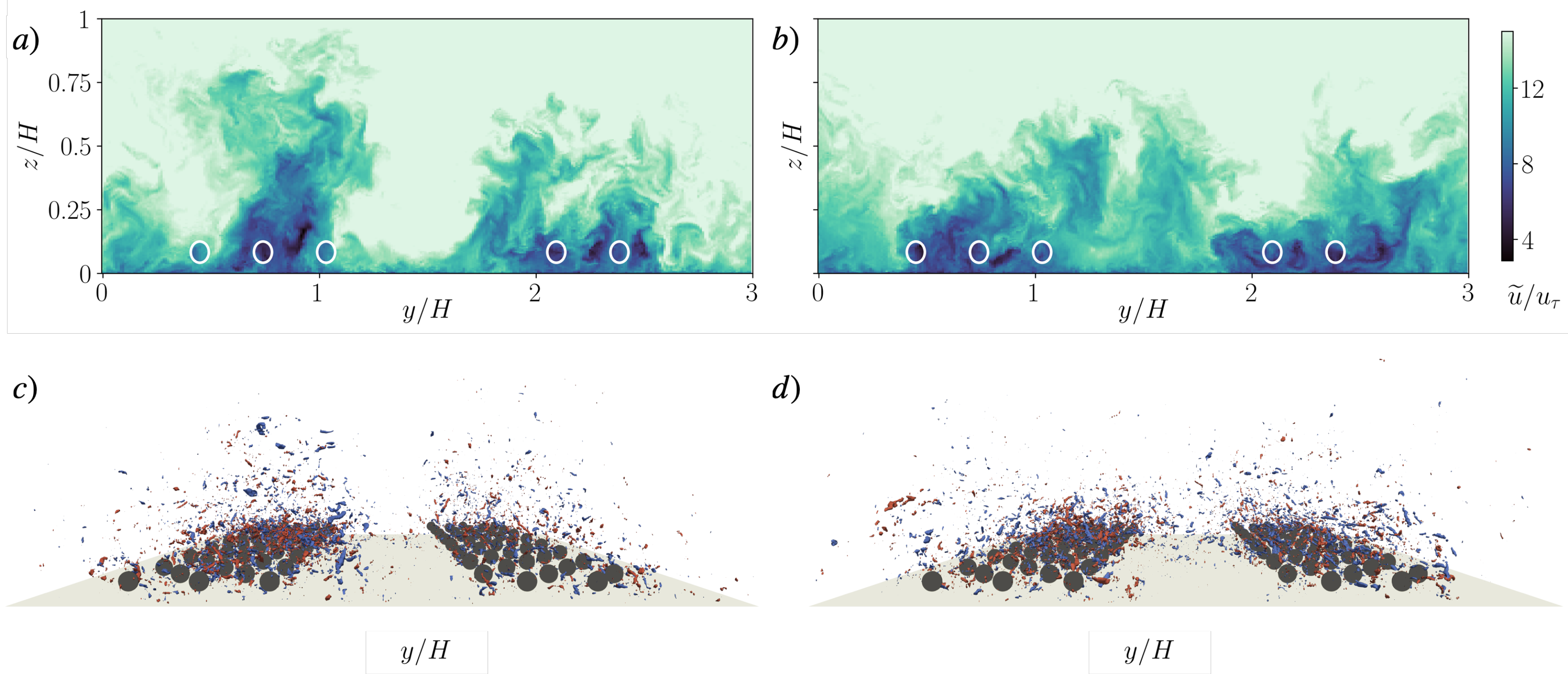}}
  \caption{(a, b) Pseudocolor plots of normalized streamwise velocity $(\tilde{u}/u_\tau)$ at different timesteps, taken at a streamwise plane intersecting the element locations for the case S3.5-5-2. White circles indicate element positions. (c, d) Visualization of streamwise vortical structures using iso-surfaces of signed swirling strength (4\% of maximum $\lambda_{ci}^2$). Panels (c) and (d) correspond to the same timesteps as panels (a) and (b), respectively.}
\label{fig:inst-reversal}
\end{figure}

Figure~\ref{fig:inst-reversal}(a, b) presents two realizations of the instantaneous streamwise velocity for this case, taken approximately 12 large eddy turnovers ($H/u_\tau$) apart. 
The color bar is capped so that the velocity values exceeding 80\% of the maximum appear in a uniform light shade, enhancing the visibility of \R{HMRs}.
In figure~\ref{fig:inst-reversal}(a), a dominant \R{HMR} is clearly visible at the center of the domain, while low-momentum fluid remains near the roughness elements and extends upward into the outer layer. 
However, in figure~\ref{fig:inst-reversal}(b), an \R{LMR} occupies the center of the domain, penetrating deep into the outer layer. 
While the low-momentum fluid is always present near the roughness elements due to the drag they impose, these two realizations illustrate that instantaneous secondary circulations can organize themselves to either advect this low-momentum fluid upward or, at a different timestep, switch circulation direction and transport it laterally at the center—despite the roughness arrangement remaining unchanged.
Thus, while an HMP appears at the domain center in the long-time-averaged visualization, this does not imply \R{a continuous presence of HMR at the center} throughout the simulation.
Instead, the \R{regions of low- and high-momentum} alternate, and what we observe in the long-time-averaged field is the superposition of these alternating structures. 
Depending on their relative strength and persistence over time, one pathway may appear dominant in the long-time-averaged visualization.

Figure~\ref{fig:inst-reversal}(c, d) shows the iso-surfaces of two-dimensional signed swirl strength, which indicates the strength of local swirling in the streamwise direction.
The swirl strength ($\lambda_{ci}$) is computed as the imaginary component of the complex eigenvalue of the two-dimensional velocity gradient tensor, computed in the spanwise-wall-normal plane, and is signed based on the direction of vorticity \citep{ZhouBalachandar1999, Wu_Christensen_2006, Anderson2015Secflow}.
The figure displays iso-surfaces corresponding to 4\% of the maximum $\lambda_{ci}^2$, a threshold selected to enhance the visualization of stronger circulations while minimizing background noise.
In figure~\ref{fig:inst-reversal}(c), most of the vorticity is concentrated near the elements, while the valley between the arrays exhibits a prominent region of low circulation. 
In contrast, figure~\ref{fig:inst-reversal}(d) shows a noticeable increase in vorticity within the valley. 
Taken together with the streamwise velocity visualizations, these observations indicate that \R{LMRs} are associated with strong streamwise circulations, whereas \R{HMRs} tend to suppress them.
This observation aligns with the findings in the literature, where it is reported that \R{LMPs} exhibit stronger random fluctuations, while \R{HMPs} remain comparatively calmer \citep{Barros2014, Anderson2015Secflow, Kevin2017, Vanderwel2019}. 
Additional analysis of instantaneous vortices (not shown) reveals that, across all cases listed in table~\ref{tab:staggered reversal}, the largest instantaneous vortex is approximately an order of magnitude smaller than the delta-scale vortex. 
This analysis also indicates that larger instantaneous circulations generally tend to be weaker, while stronger circulations are typically confined to smaller spatial regions.
These findings confirm that delta-scale vortices do not exist instantaneously but instead arise as a consequence of time-averaging.

The observations presented in this subsection give rise to three key questions:
\begin{enumerate}
\item For how long does a given \R{region of low- or high-momentum} persist at the center? 
\item How frequently do the \R{regions} switch between one another?
\item Is this behavior universal, or does it depend on the roughness arrangement?
\end{enumerate}
Exploring these questions is essential to understanding why some cases exhibit symmetrical secondary flows, while others appear disrupted. The next subsection investigates these aspects in greater detail, shedding light on the underlying flow dynamics.

\subsection{Conditional averaging of instantaneous flow field} \label{sec: conditional-average}
In this subsection, we investigate the conditionally averaged flow fields for the cases listed in table~\ref{tab:staggered reversal} and table~\ref{tab:disproving sl/D}.
As discussed in the previous subsection, the \R{HMR} and the \R{LMR} can both occur at the domain center at different instances, indicating reversals in the swirling direction of the secondary flows.
Since downdrafts and updrafts serve as key indicators for the presence of \R{HMRs} and \R{LMRs}, respectively, we categorize the instantaneous flow into two distinct events, based on the relative occurrence of updrafts and downdrafts at the domain center.
For this conditional averaging, velocity data from eight spanwise locations are considered: four located closest to the center of the domain, and two each near the edges in the spanwise direction. 
These edge locations also represent valleys between elements due to the periodic boundary condition. 
At each of these locations, vertical velocity values from the entire streamwise and vertical extent are included, resulting in approximately 885k grid points at which updraft and downdraft occurrences are evaluated for each timestep.
Consequently, two distinct averaged velocity fields are generated: one representing the event when updrafts outnumber downdrafts, and the other corresponding to the event when downdrafts predominate.

\R{To formalize this classification, let $w(\boldsymbol{x}_i,t)$ be the instantaneous vertical velocity at grid point $\boldsymbol{x}_i$ and time $t$, with $\mathcal{D}$ denoting the set of diagnostic grid points.
At each time instant $t$, the number of updraft and downdraft points is computed as}

\begin{equation}
\begin{aligned}
N_{\text{up}}(t)   &= \# \{ \boldsymbol{x}_i \in \mathcal{D} \, | \, w(\boldsymbol{x}_i,t) > 0 \}, \\
N_{\text{down}}(t) &= \# \{ \boldsymbol{x}_i \in \mathcal{D} \, | \, w(\boldsymbol{x}_i,t) \leq 0 \}.
\end{aligned}
\end{equation}

\noindent \R{The flow field at time $t$ is classified as updraft‑dominated if $N_{\text{up}}(t) > N_{\text{down}}(t)$ and as downdraft‑dominated otherwise. 
The corresponding conditional averages are then defined as}
\begin{equation}
\begin{aligned}
\overline{u}_{i_{c+}}(\boldsymbol{x}) &= \frac{1}{T_{\text{up}}} 
\sum_{\substack{t \, : \, N_{\text{up}}(t) > N_{\text{down}}(t)}} 
\tilde{u}_i(\boldsymbol{x},t)\ , \\
\overline{u}_{i_{c-}}(\boldsymbol{x}) &= \frac{1}{T_{\text{down}}} 
\sum_{\substack{t \, : \, N_{\text{down}}(t) \ge N_{\text{up}}(t)}} 
\tilde{u}_i(\boldsymbol{x},t)\ ,
\end{aligned}
\end{equation}
\noindent \R{where $T_{\text{up}}$ and $T_{\text{down}}$ are the number of time instants satisfying each condition.}

\begin{figure}
  \centerline{\includegraphics[width=\textwidth]{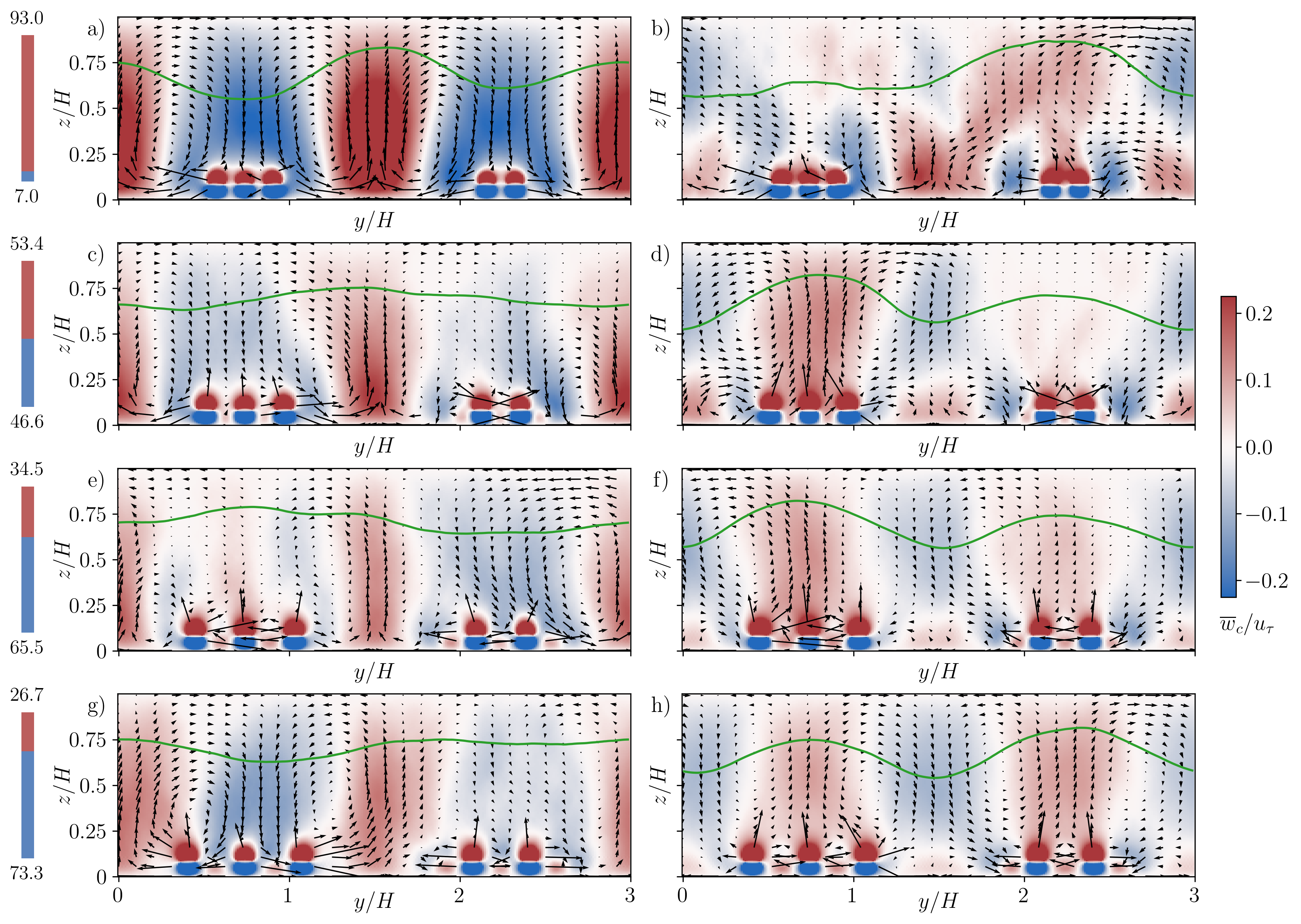}}
  \caption{Conditionally averaged pseudocolor plots of vertical velocity at a streamwise location aligned with the elements for the cases listed in table~\ref{tab:staggered reversal}: (a, b) S2-5-2, (c, d) S2.75-5-2, (e, f) S3.5-5-2, and (g, h) S4-5-2. The left column corresponds to events where updrafts outnumber downdrafts, while the right column represents events dominated by downdrafts. The vertical bar on the left side of each row indicates the percentage of time updrafts were dominant (red, with numerical value at the top) and the percentage of time downdrafts were dominant (blue, with numerical value at the bottom). Black arrows indicate the vectors of spanwise and vertical velocity. Green line indicates contour of 95\% of maximum mean streamwise velocity.}
\label{fig:conditional averaging}
\end{figure}

Figure~\ref{fig:conditional averaging} presents the conditionally averaged vertical velocity fields for the cases listed in table~\ref{tab:staggered reversal}. 
The left column, i.e., figure~\ref{fig:conditional averaging}(a, c, e, g), corresponds to events where updrafts outnumber downdrafts at the domain center, while the right column, i.e., figure~\ref{fig:conditional averaging}(b, d, f, h), depicts events dominated by downdrafts.
The vertical bar on the left side of each row indicates the percentage of simulation time for which each specific event occurs, with the extent of red color denoting dominant updrafts and blue representing dominant downdrafts.
It is evident that in all cases—including those where symmetrical secondary flows with a dominant polarity were observed—the \R{updrafts} and \R{downdrafts} do not persist for the entire duration of the simulation. 
For instance, in case S2-5-2, an updraft exists at the domain center for 93\% of the simulation time, while downdrafts dominate 7\% of the time. 
As a result, the conditionally averaged field shown in figure~\ref{fig:conditional averaging}(a) closely resembles the long-time averaged field in figure~\ref{fig:mean_vort_2-3clubbed}(a), as the events with opposite polarity lack sufficient temporal persistence to leave a significant imprint on the long-time averaged flow field.
As the parameter $s_a/H$ is decreased, moving to the case S2.75-5-2, events with dominant downdrafts become substantially more frequent, constituting 46.6\% of the simulation timesteps, while updrafts persist for the remaining 53.4\% of the time. 
In periods where downdrafts dominate, the streamwise velocity clearly exhibits strong spanwise modulation, with HMPs appearing at the domain center and LMPs aligned over the roughness elements. 
In contrast, when updrafts dominate, the modulation is notably weaker, with weak LMPs near the domain center and weak HMPs near the elements.
This observation is significant because, in the long-time averaged field shown in figure~\ref{fig:mean_vort_2-3clubbed}(b), an HMP appears at the domain center, despite the updraft (and its associated LMP) persisting longer. 
This highlights that the long-time average may not always reflect the flow structures that are most frequently present throughout the simulation.
Another interesting observation for this case is the reversal of flow pathways at the roughness elements.
Specifically, figure~\ref{fig:conditional averaging}(c) shows downdrafts clearly aligned with the elements, while figure~\ref{fig:conditional averaging}(d) shows updrafts aligned over these arrays. 
This behavior differs from the findings of \citet{Vanderwel2019}, who reported that for flow over streamwise-aligned ridges, \R{LMRs} remained anchored to elevated surface regions, extending deep into the boundary layer as narrow and meandering streaks. 
We attribute this discrepancy to differences in the physical nature of roughness elements and the presence of significant streamwise gap in our roughness configuration.
Overall, despite both the conditionally averaged flow fields showcasing the presence of delta scale vortices, as no event has dominant presence throughout the simulation, the long-time averaged visualization shows seemingly disrupted secondary flows.
Further decreasing $s_a/H$ results in case S3.5-5-2, where downdrafts become even more frequent, dominating approximately 65.5\% of the simulation duration compared to 34.5\% for updrafts.
Similar to the previous case, downdraft-dominated events strongly modulate streamwise velocity, while updraft-dominated events exhibit only weak modulation.
Interestingly, all delta-scale vortices observed in the conditional averages for this case rotate in a counter-clockwise direction. 
This explains why a prominent counter-clockwise delta-scale vortex appears clearly in the long-time averaged visualization (figure~\ref{fig:mean_vort_2-3clubbed}(c)), even though the rest of the flow appears somewhat disrupted due to the lack of persistent dominance of one particular pathway.
This is likely due to the dominance of one-sided large-scale rotational motions in this configuration, a feature previously identified by \citet{Kevin2017}.
Finally, in case S4-5-2, where clear delta-scale secondary flows and dominant pathways appear in the long-time averaged visualization, downdrafts dominate at the domain center for approximately 73\% of the simulation.
However, the reversed secondary flow pattern remains less symmetric than in case S3.5-7-2 (figure~\ref{fig:mean_vort_1-4clubbed}(f)), where downdrafts dominate approximately 95\% of the simulation time, resulting in a symmetrical long-time averaged flow pattern. 
Overall, as $s_a/H$ is progressively decreased, we observe a consistent shift towards downdraft dominance at the domain center. 
The configurations transition from exhibiting a strong central updraft to a pronounced central downdraft, eventually reaching a level where this reversal is clearly evident in the long-time-averaged flow field.

\begin{figure}
  \centerline{\includegraphics[width=0.875\textwidth]{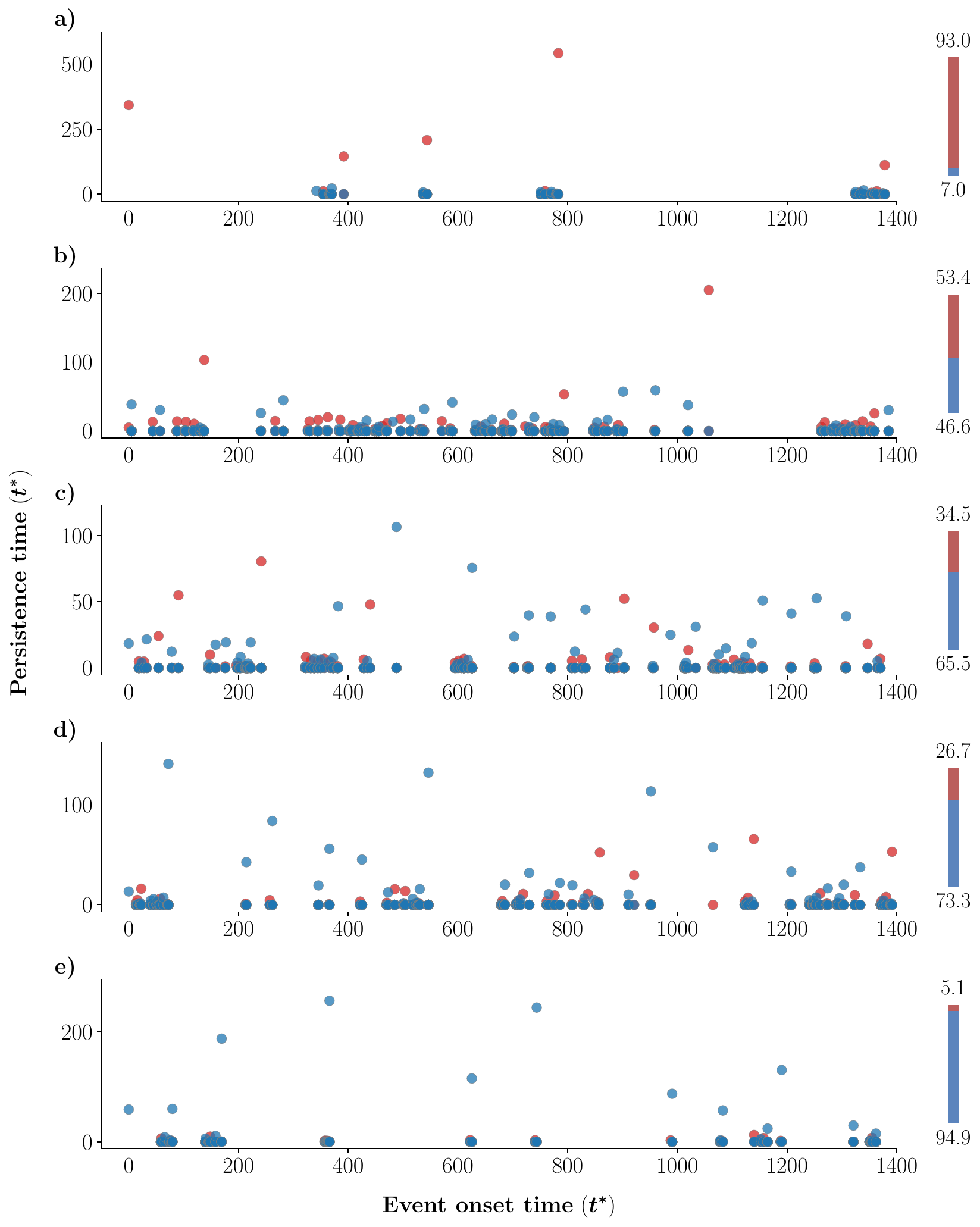}}
  \caption{Temporal persistence of \R{updraft and downdraft events} for cases: a) S2-5-2, b) S2.75-5-2, c) S3.5-5-2, d) S4-5-2, e) S3.5-7-2. Red color denotes a net updraft at the domain center, while blue color denotes a net downdraft. The vertical bar on the right side of each row indicates the percentage of time updrafts were dominant (red, with numerical value at the top) and the percentage of time downdrafts were dominant (blue, with numerical value at the bottom).}
\label{fig:pathway time switch}
\end{figure}

From the analysis so far, it is abundantly clear that for most of the secondary flow simulations, the \R{updrafts} and \R{downdrafts, and the associated LMRs and HMRs,} do not remain stable and may interchange with one another. 
However, it remains unclear whether these reversals occur in a periodic manner or follow a more chaotic, non-periodic pattern.
To investigate this, figure~\ref{fig:pathway time switch} marks the starting timestep where a particular \R{updraft or downdraft event} occurs and the marker's extent on y axis indicates the duration for which that \R{event} persists.
The analysis is conducted for the cases mentioned in table~\ref{tab:staggered reversal} along with the case S3.5-7-2 in table~\ref{tab:disproving sl/D} to emphasize dynamics of a case where downdrafts dominate at the center of the domain for the majority of the simulation.

For case S2-5-2, we observe that the \R{updrafts} at the domain center are relatively stable and can persist continuously for extended durations. 
In contrast, \R{downdrafts} occur infrequently and in a non-periodic manner, and they are typically short-lived.
For the intermediate cases shown in figure~\ref{fig:pathway time switch}(b, c), the average temporal extent of both \R{updrafts} and \R{downdrafts} is significantly reduced. 
The vertical axis scale is adjusted individually for each subplot in the figure to enhance visual clarity. 
In these cases, the relative occurrence of \R{updrafts} and \R{downdrafts} at the domain center appears roughly balanced, with the exception of two long-duration \R{updrafts} events in case S2.75-5-2.
Due to the reduced persistence of individual \R{events}, these intermediate configurations undergo frequent, non-periodic reversals. 
As a result, the secondary flow patterns appear disrupted in the long-time-averaged visualizations.
As $s_a/H$ decreases further, in cases such as S4-5-2 and S3.5-7-2, we find that \R{downdrafts} at the domain center begin to stabilize and persist for longer durations compared to \R{updrafts}.
With \R{downdrafts} sustaining for extended periods, reversals occur less frequently, and the long-time-averaged flow fields exhibit greater symmetry, with clearly organized \R{LMPs} and \R{HMPs}.

\begin{figure}
  \centerline{\includegraphics[width=1.05\textwidth]{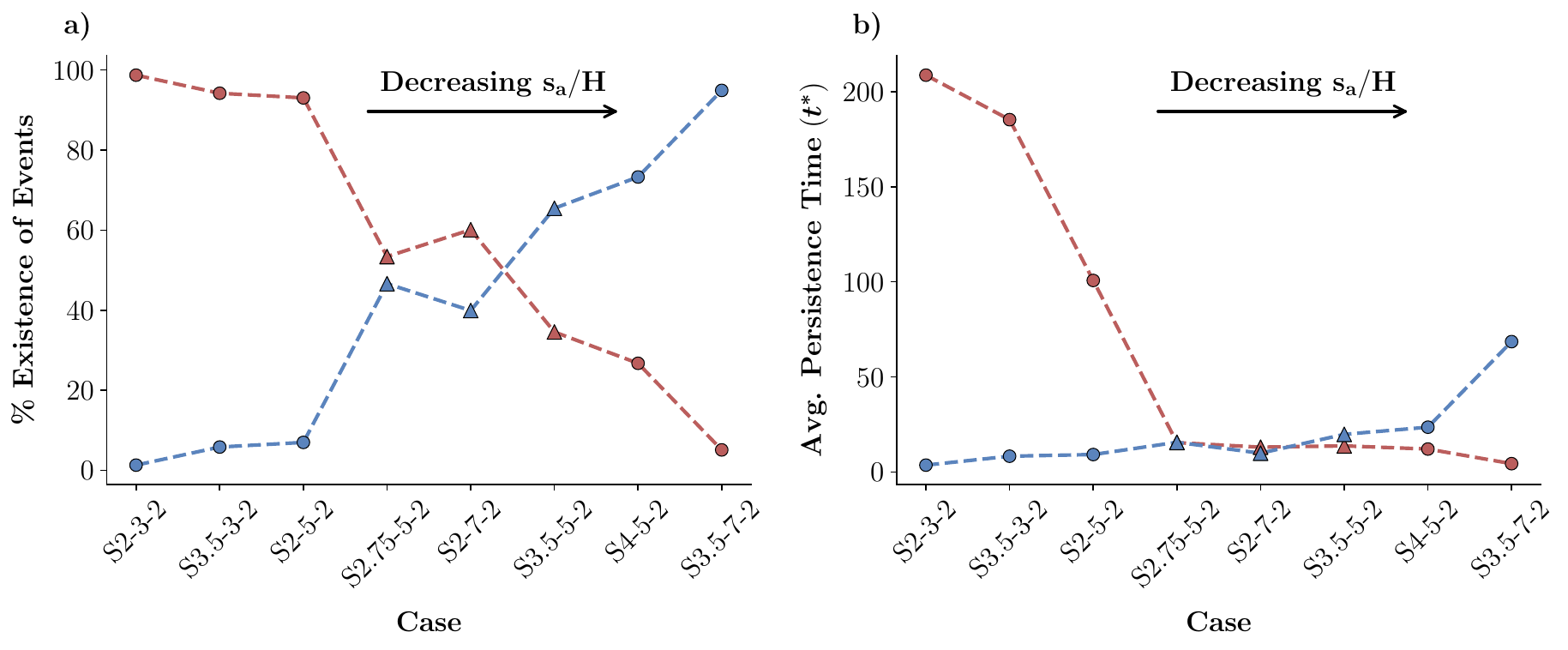}}
  \caption{(a) Percentage occurrence of \R{updraft and downdraft events} and (b) average persistence time for the cases outlined in table~\ref{tab:staggered reversal} and table~\ref{tab:disproving sl/D}. Red color denotes a net updraft at the domain center, while blue color denotes a net downdraft. Circles indicate cases exhibiting a dominant polarity in long-time-average visualization, whereas triangles indicate cases with disrupted secondary flows in long-time-average visualization. The cases are arranged in decreasing order of $s_a/H$.}
\label{fig:percentage polarity}
\end{figure}

Figure~\ref{fig:percentage polarity} provides a quantitative assessment of the visual observations discussed above. 
This figure includes all cases listed in table~\ref{tab:staggered reversal} and table~\ref{tab:disproving sl/D}, arranged in decreasing order of $s_a/H$.
The percentage existence of \R{updraft and downdraft events} refers to the fraction of the simulation during which a particular \R{event} is dominant. 
The average time of existence denotes the mean duration for which a given \R{event} persists continuously. 
To avoid the influence of very short-lived events on the average time of existence, only \R{the events} persisting for $t^*>1$ are considered.
As shown in figure~\ref{fig:percentage polarity}(a), cases in which one \R{event} exists for more than 2.5 times the duration of the other tend to exhibit a dominant polarity in the long-time averaged flow field, accompanied by clear delta-scale vortices. 
In contrast, intermediate cases show roughly equal presence of both polarities, with no single \R{event} significantly dominating the other.
A similar trend is observed in the average time of existence. 
For the intermediate cases, both \R{events} persist for nearly equal durations, whereas in cases with a dominant polarity, the dominant \R{event} lasts at least twice as long as the non-dominant one.

In conclusion, the analysis presented in this subsection reveals a clear transitional behavior as the parameter $s_a/H$ decreases from case S2-5-2 to case S3.5-7-2.
For the cases outlined in table~\ref{tab:staggered reversal} and table~\ref{tab:disproving sl/D}, it is observed that when $s_a/H \gtrsim 1.2$, delta-scale secondary flows emerge with LMPs \R{and associated updrafts} aligned at the domain center. 
These \R{updrafts} persist for the majority of the simulation time, and although reversals can occur, they are brief and contribute negligibly to the long-time averaged flow field.
When the cases lie in the regime $1.2 \gtrsim s_a/H \gtrsim 1$, the \R{updrafts} at the domain center become increasingly unstable and do not sustain for longer periods during the simulations.
As a result, the reversal occurs much more frequently, and both \R{events} appear to be short-lived.
As no particular \R{event} significantly dominates the other in this regime, the secondary flows appear to be disrupted in the long-time average visualization.
However, for these cases, delta-scale vortices can still be observed in the conditionally averaged flow field, as shown in figure~\ref{fig:conditional averaging}(c, d). 
When $s_a/H$ is reduced further such that the cases lie in the regime $s_a/H \lesssim 1$, the HMPs \R{and associated downdrafts} at the domain center begin to dominate and sustain for longer durations.
As a result, the propensity of the system for \R{reversals} decreases, and we again observe delta scale secondary flows with a prominent alignment of pathways in the long-time average visualization.

\subsection{Characterization of instantaneous vertical velocity distribution}\label{sec: PDF analysis}
In the previous subsection, we classified events based on the sign of the vertical velocity at grid points located at the domain center, thereby distinguishing between updrafts and downdrafts. 
While this approach provided valuable insight into the switching behavior between \R{these events} at the domain center, it did not capture variations in the magnitude of vertical velocity within individual events. 
As observed in figure~\ref{fig:pathway time switch}, different roughness configurations influence the stability of the \R{events} sustained at the domain center. 
This raises the question of whether the time series of vertical velocity exhibits distinct signatures depending on whether a dominant event—either updraft or downdraft—is sustained at the domain center, or whether the flow frequently switches between the two.
Therefore, in this subsection, we analyze the temporal evolution of vertical velocity to gain a deeper understanding of its dynamics during updraft and downdraft events.

\begin{figure}
  \centerline{\includegraphics[width=1.1\textwidth]{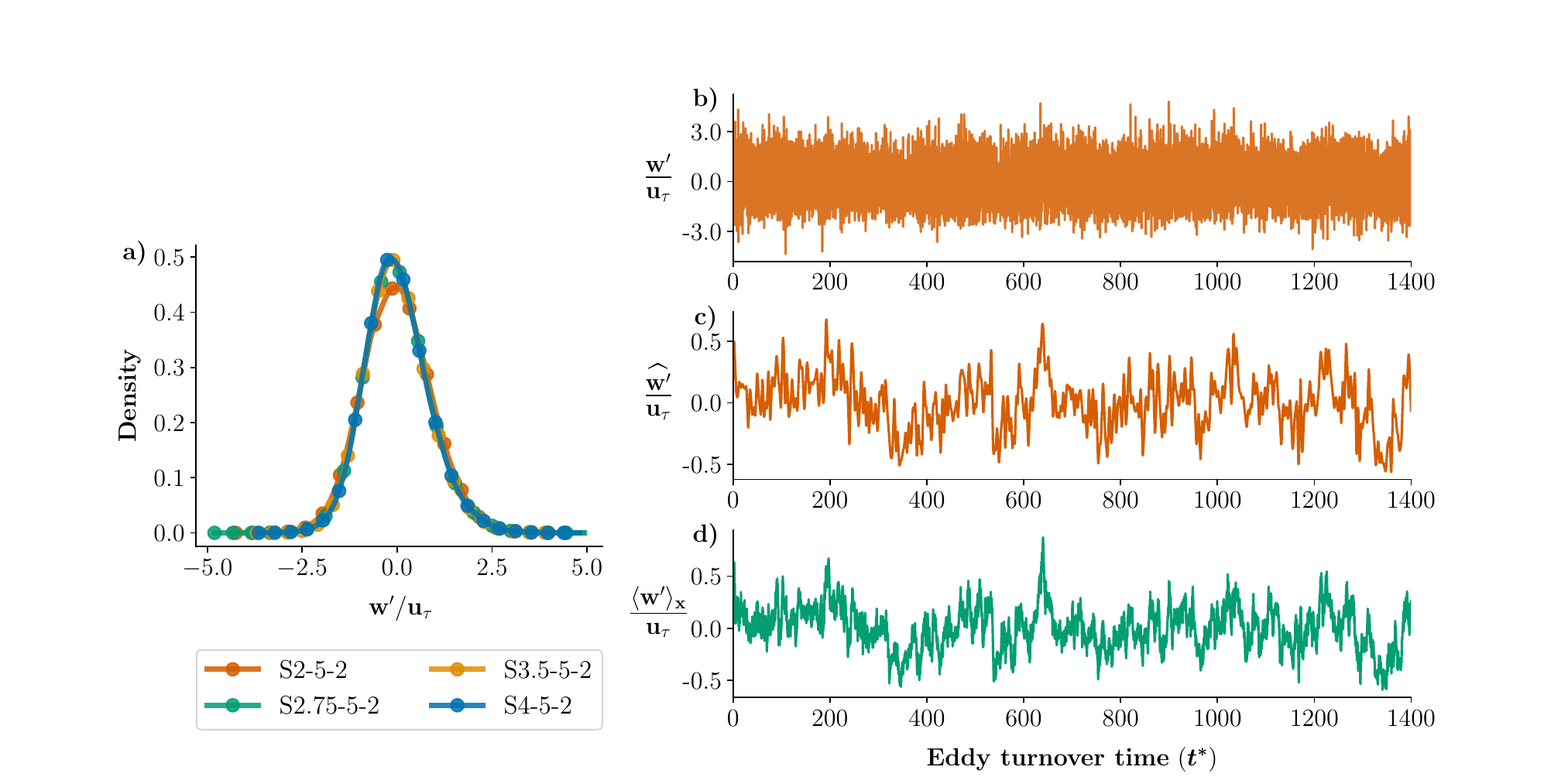}}
  \caption{(a) Probability density functions of the time series of vertical velocity fluctuations for the cases listed in table~\ref{tab:staggered reversal}. 
  Data are sampled at two streamwise positions: the location of the first row of elements (solid lines), and the midpoint between the first and second element rows (circles). 
  In both cases, the spanwise and vertical location of the sampling point corresponds to the center of the Y–Z plane, i.e., $y = L_y/2$ and $z = H/2$.
  (b) Time series of vertical velocity fluctuations for case S2-5-2, sampled at a grid point aligned with the first element row. (c) Low-pass filtered version of the signal shown in (b). (d) Signal sampled and averaged over every $x$ location at $y = L_y/2$ and $z = H/2$.}
\label{fig:pointwise signal and PDF}
\end{figure}

Figure~\ref{fig:pointwise signal and PDF}(a) presents the probability density functions (PDFs) of the time series of vertical velocity fluctuations for the cases listed in table~\ref{tab:staggered reversal}. 
To assess the influence of streamwise heterogeneity introduced by the roughness arrangement, vertical velocity time signals are sampled at two streamwise locations: one coinciding with the first row of elements, and the other situated at the midpoint between the first and second element rows. 
The spanwise position of the sampling point is fixed at the domain center (i.e., $y = L_y/2$), and the vertical position is located at half the half-channel height (i.e., $z = H/2$).
As seen from the figure, the PDFs at the two streamwise locations exhibit a high degree of similarity, suggesting that the streamwise heterogeneity in the roughness layout does not significantly affect the statistical properties of vertical velocity at this particular spanwise location and height.

In this figure, the case S2-5-2 exhibits a distinctly broader PDF—indicating a higher standard deviation—while the PDFs for the other cases appear remarkably similar.
The similarity between the intermediate cases (S2.75-5-2 and S3.5-5-2) and S4-5-2 is intriguing, as it suggests that the frequent pathway reversals characterizing the intermediate cases do not leave a distinct imprint on the temporal statistics of vertical velocity, compared to cases with a persistent central pathway and infrequent reversals. 
Instead, the broader distribution observed for S2-5-2, along with the monotonic (albeit modest) decrease in standard deviation across the other cases, implies that the intensity of fluctuations is primarily governed by how long the flow locally maintains an updraft—typically associated with stronger turbulent activity—or a downdraft, which tends to exhibit weaker fluctuations.

Figure~\ref{fig:pointwise signal and PDF}(b) shows the time series of vertical velocity fluctuations for case S2-5-2. 
While the signal exhibits strong turbulent fluctuations, it also reveals an underlying low-frequency modulation. 
To isolate and examine this low-frequency behavior, we apply the maximal overlap discrete wavelet transform (MODWT) using a Coiflet2 wavelet. 
The MODWT smooth component acts as a zero-phase, shift-invariant low-pass filter, preserving the temporal alignment of the low-frequency content with the original signal \citep{Percival1997}.
The support of the scaling function used for filtering spans approximately 45{,}000 timesteps ($t^* = tu_\tau/h \approx 9.1$), applied to a signal with a total duration of 7{,}000{,}000 timesteps ($t^* = 1417.5$).
The resulting low-pass filtered signal is shown in figure~\ref{fig:pointwise signal and PDF}(c).
Figure~\ref{fig:pointwise signal and PDF}(d) presents the streamwise-averaged vertical velocity fluctuations, obtained by averaging the signal across all the streamwise grid points at $y = L_y/2$ and $z = H/2$ at each timestep. 
The resulting time series closely resembles the low-pass filtered signal derived from a single point, indicating that the streamwise averaging operation at this spanwise and vertical location effectively preserves the long-time variations in the vertical velocity fluctuations observed during the simulation.
Therefore, to mitigate sensitivity to sampling location, we hereafter focus on the streamwise-averaged vertical velocity signal.

\begin{figure}
  \centerline{\includegraphics[width=0.925\textwidth]{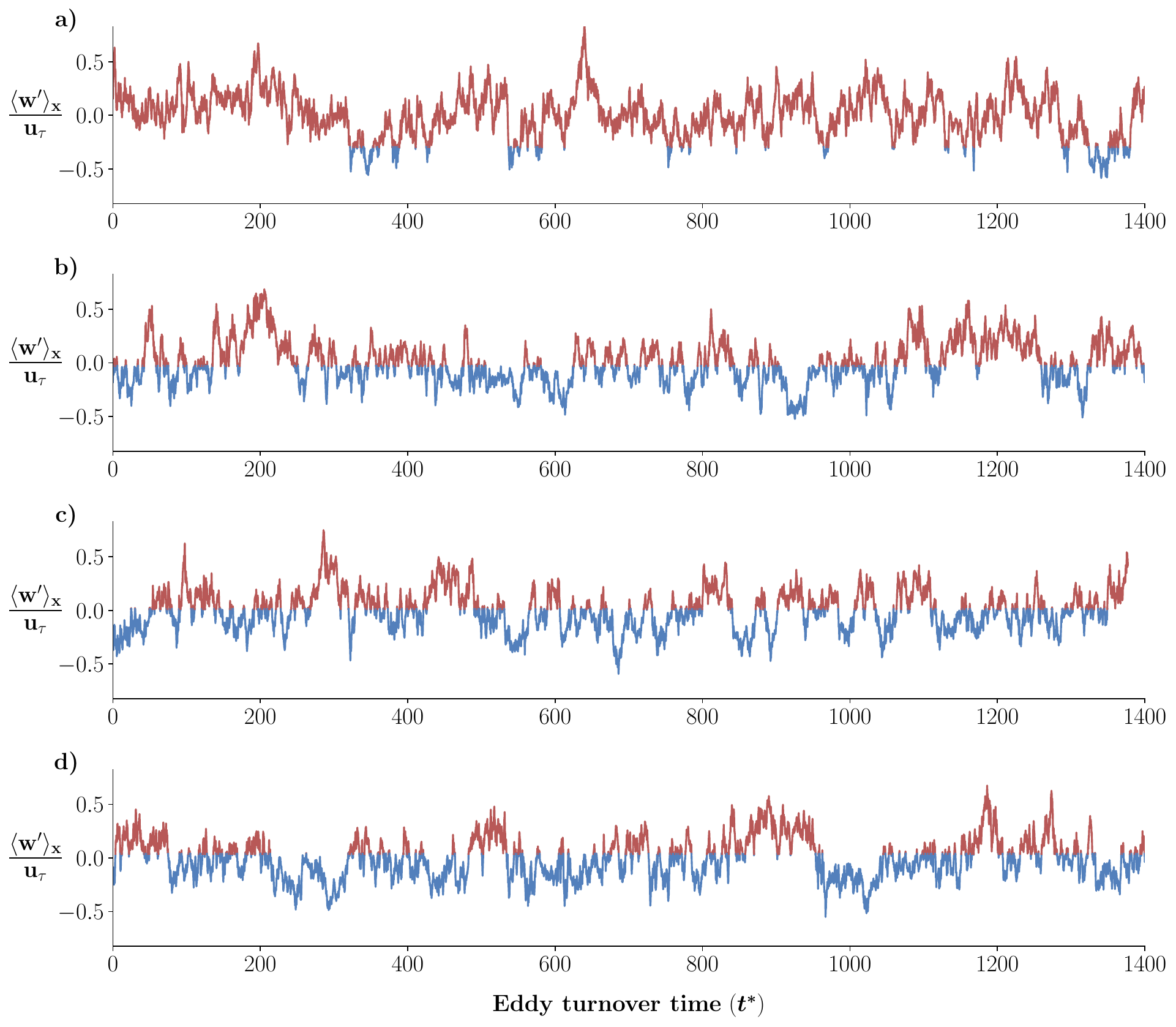}}
  \caption{Time series of vertical velocity fluctuations that are sampled and averaged over every $x$ location at $y = L_y/2$ and $z = H/2$, for cases: a) S2-5-2, b) S2.75-5-2, c) S3.5-5-2, d) S4-5-2. Red color denotes a positive vertical velocity $(\langle \overline{w} \rangle_x + \langle w^\prime \rangle_x > 0)$ while blue color denotes a negative vertical velocity. }
\label{fig:w signal xaveraged}
\end{figure}

Figure~\ref{fig:w signal xaveraged} shows the streamwise-averaged vertical velocity fluctuations at $y = L_y/2$ and $z = H/2$ for the cases listed in table~\ref{tab:staggered reversal}.
To enable direct comparison with figure~\ref{fig:pathway time switch}, which illustrates the time-dependent switching of \R{updraft and downdraft events}, the signal is color-coded such that red indicates regions where the vertical velocity is positive (i.e., $\langle \overline{w} \rangle_x + \langle w' \rangle_x > 0$), while blue indicates regions where it is negative.
From the figure, it is evident that all four cases exhibit similar low-frequency, periodic modulations in the vertical velocity, with the standard deviation remaining within 10\% across all cases.
Although conditional averaging previously indicated that the \R{updraft} at the domain center is relatively stable in case S2-5-2, persisting for long durations with infrequent reversals, this stability of the \R{event} does not necessarily correspond to a vertical velocity signal with lower temporal variability.
As seen in figure~\ref{fig:w signal xaveraged}(a), the vertical velocity remains predominantly positive for extended periods, but its magnitude continues to undergo quasi-periodic oscillations, similar to those observed in the other cases (figures~\ref{fig:w signal xaveraged}(b--d)).
However, because these oscillations mostly occur within the positive regime, they do not manifest as reversals in the conditionally averaged statistics.
In contrast, for the intermediate cases (S2.75-5-2 and S3.5-5-2), the vertical velocity oscillations frequently cross zero, leading to repeated sign changes.
This is primarily due to the mean vertical velocity at this location being close to zero.
Thus, the more frequent switching observed in figure~\ref{fig:pathway time switch}(c), relative to figure~\ref{fig:pathway time switch}(a), reflects the near-neutral mean state rather than an inherently more chaotic fluctuation pattern.
In fact, as noted earlier in the discussion of figure~\ref{fig:pointwise signal and PDF}, updraft-dominated regions are associated with stronger turbulent fluctuations. Therefore, case S2-5-2, which sustains more persistent updrafts, exhibits higher overall variance in vertical velocity compared to the intermediate cases, despite appearing less dynamic in terms of switching \R{between updrafts and downdrafts}.

\begin{figure}
  \centerline{\includegraphics[width=0.875\textwidth]{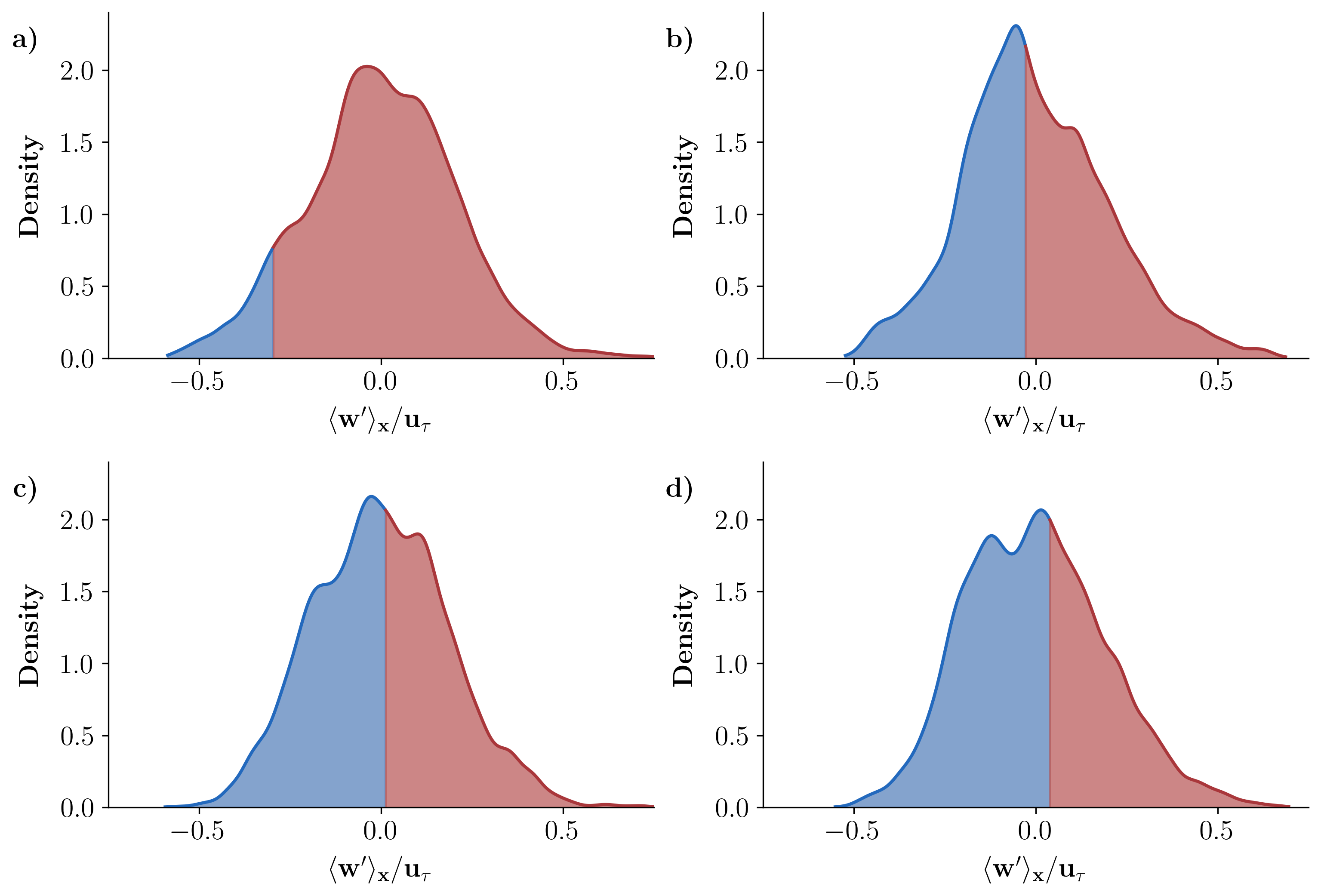}}
  \caption{Probablity density function of time series of vertical velocity fluctuations that are sampled and averaged over every $x$ location at $y = L_y/2$ and $z = H/2$, for cases: a) S2-5-2, b) S2.75-5-2, c) S3.5-5-2, d) S4-5-2. Red
  region denotes a positive vertical velocity $(\langle \overline{w} \rangle_x + \langle w^\prime \rangle_x > 0)$ while blue region denotes a negative vertical velocity.}
\label{fig:PDF w signal xaveraged}
\end{figure}

Figure~\ref{fig:PDF w signal xaveraged} shows the PDFs of the streamwise-averaged vertical velocity fluctuations discussed above.
The PDFs are color-coded such that red corresponds to regions where the vertical velocity is positive (i.e., $\langle \overline{w} \rangle_x + \langle w^\prime \rangle_x > 0$), and blue corresponds to negative vertical velocity.
In contrast to the unimodal distributions observed in figure~\ref{fig:pointwise signal and PDF}, the streamwise-averaged PDFs in this figure reveal a clear multi-modal character across all cases.
For cases S2-5-2 and S4-5-2, which exhibit a dominant polarity in the long-time-averaged flow field, the prominent modes lie predominantly on one side of the mean: positive for S2-5-2 and negative for S4-5-2. 
However, for the intermediate cases (S2.75-5-2 and S3.5-5-2), the modes are distributed on both sides of the mean, reflecting frequent transitions between updraft- and downdraft-dominated states.
These observations imply that secondary circulations are inherently unsteady, regardless of whether the time-averaged flow exhibits a persistent polarity or not. 
In all cases, the vertical velocity signal fluctuates between multiple preferred states, as evidenced by the modal structure of the PDFs. 
This behavior persists even in flows that do not visibly exhibit \R{frequent} switching, indicating that temporal variability in the vertical momentum transport is a fundamental feature of the secondary flow dynamics.

\section{Conclusion} \label{sec: Conclusion}
In this study, we conducted a series of LES to examine the behavior of secondary flows induced by multi-column roughness configurations.
Each column consisted of multiple, individually resolved roughness elements arranged in both the streamwise and spanwise directions. 
Wind turbines were used as the roughness elements to represent flow through porous media.
The local spanwise gap between the elements as well as the number of element columns were systematically varied to construct a range of roughness configurations. 
Subsequently, we analyzed how these geometric modifications influenced the polarity and temporal dynamics of the resulting secondary flows.

The study first examined the time-averaged characteristics of secondary flows in \S\ref{sec: Time averaged results}. 
The key findings from this section are summarized below:
\begin{enumerate}
    \item The analysis established that the parameter $s_a/H$—defined as the ratio of the spanwise gap between roughness elements in adjacent columns to the boundary layer height—plays a critical role in determining the polarity of secondary flows in multi-column roughness configurations.
    \item Using the energy tube framework introduced by \citet{Meyers2013}, we demonstrated how variations in $s_a/H$ influence the entrainment pathways of the \textit{mke}. 
    The geometry of these energy tubes was shown to encode clear information about the streamwise circulation patterns associated with secondary flows. 
    This methodology offers a valuable alternative to the traditional \textit{tke} budget analysis for secondary flows, particularly in complex roughness arrangements where symmetry-based simplifications in the \textit{tke} budget are not applicable.
    \item Based on the value of $s_a/H$, the cases in this study can be categorized into three distinct regimes.
    When $s_a/H \gtrsim 1.2$, the long-time averaged flow field showed delta-scale secondary flows with HMPs and the corresponding downdrafts aligned with the elements, and LMPs and associated updrafts aligned with the valley between the elements.
    When $1.2 \gtrsim s_a/H \gtrsim 1$, the secondary flow structures in the long-time-averaged field appear disrupted, lacking the presence of symmetrical delta-scale vortices.
    For $s_a/H \lesssim 1$, the long-time-averaged field once again shows delta-scale secondary flows, but with a polarity reversal: LMPs and associated updrafts are now aligned with the element columns, while HMPs and associated downdrafts appear over the valleys.
\end{enumerate}

To uncover why different $s_a/H$ regimes give rise to either coherent or disrupted secondary flows in the long-time-averaged fields, we examined the instantaneous flow behaviour in \S\ref{sec: Instantaneous results}, focusing on the transient mechanisms that are obscured by time averaging.
The key findings from this section are summarized below:
\begin{enumerate}
    \item Across all the cases, regardless of the $s_a/H$ regime, the \R{updrafts and downdrafts, along with their associated LMRs and HMRs, respectively,} do not remain permanently aligned with the roughness topography. 
    Instead, they alternate between each other in a chaotic, non-periodic manner over time.
    \item In the regimes $s_a/H \gtrsim 1.2$ and $s_a/H \lesssim 1$, where a dominant polarity of secondary flow is observed in the long-time-averaged visualizations, the dominant \R{vertical motion} tends to persist for the majority of the simulation. 
    While reversals may still occur, they are generally infrequent and short-lived.
    In contrast, in the intermediate regime ($1.2 \gtrsim s_a/H \gtrsim 1$), where the secondary flow structures appear disrupted in the long-time-averaged field, neither \R{vertical event} dominates over time.
    Instead, frequent reversals occur, and both \R{the updraft and downdraft events} tend to persist for similar durations, indicating that the disrupted appearance arises due to the comparable strength and temporal prevalence of the competing \R{vertical motions}.
    Nonetheless, delta-scale secondary structures with opposite rotational directions can be observed for these cases in the conditionally averaged flow field.
    \item Analysis of the vertical velocity time series showed that frequent \R{switching between the vertical events} does not translate into a qualitatively different temporal signal.
    Even in cases with a persistent \R{event}, the low-frequency signal of vertical velocity exhibits similar quasi-periodic oscillations to those seen in cases with frequent reversals. 
    This showed that the increased frequency of \R{switching between the events} reflects a near-zero mean vertical velocity rather than an inherently more chaotic fluctuation pattern.
    \item The PDFs of streamwise-averaged vertical velocity exhibited a clear multi-modal structure across all cases, indicating that the secondary circulations are inherently unsteady and that the vertical velocity fluctuates between multiple preferred states. 
    In cases with a dominant polarity, these preferred states share the same sign of vertical velocity, while in intermediate cases, they span both positive and negative values, indicating that neither \R{event} dominates and both play an important role in shaping the flow dynamics.
    \item Overall, the analysis in this section showed that the temporal variability in the vertical momentum transport is a fundamental feature of the secondary flow dynamics.
\end{enumerate}

Taken together, these findings provide a comprehensive picture of how the geometric arrangement of roughness elements governs not only the mean structure but also the temporal behavior of secondary flows.
The specific $s_a/H$ regimes and the values of the length scales reported in \S\ref{sec: Time averaged results} are expected to be sensitive to the specific geometry and physical characteristics of the roughness elements, as these directly influence the \textit{mke} entrainment pathways and, in turn, the global polarity of the secondary flows. 
By contrast, the insights into the instantaneous structure and temporal dynamics appear to reflect more general features of roughness-induced secondary flows. 
The dynamical behaviours documented in this study are consistent with the trends observed in prior investigations conducted over markedly different surface types \citep{Kevin2017, Vanderwel2019, Anderson2019, Wangsawijaya2020}.

\backsection[Acknowledgements]{This work used the Anvil supercomputer at Purdue University through allocation ATM180022 from the Advanced Cyberinfrastructure Coordination Ecosystem: Services \& Support (ACCESS) program, which is supported by National Science Foundation grants \#2138259, \#2138286, \#2138307, \#2137603, and \#2138296.
The authors also acknowledge the Texas Advanced Computing Center (TACC) at The University of Texas at Austin for providing high performance computing resources that have contributed to the research results reported within this paper.}

\backsection[Funding]{This material is based upon work supported by, or in part by, the Army Research Laboratory and the Army Research Office under grant number W911NF-22-1-0178.}

\backsection[Declaration of interests]{The authors report no conflict of interest.}

\bibliographystyle{jfm}

\bibliography{bibliography.bib}

\end{document}